\documentclass[preprint,12pt,authoryear]{elsarticle}

\usepackage[a4paper, top=2.5cm, bottom=2.5cm, left=1.8cm, right=1.7cm]{geometry}
\usepackage{amsmath}
\usepackage{cleveref}

\usepackage{xcolor}

\usepackage{amssymb}
\usepackage{subfig}

\journal{Elsevier}

\begin{document}

\begin{frontmatter}

\title{Bulk and fracture process zone contribution to the rate-dependent adhesion amplification in viscoelastic broad-band materials}

\author[inst1]{Ali Maghami}
\author[inst2]{Qingao Wang}
\author[inst1]{Michele Tricarico}
\author[inst1,inst3]{Michele Ciavarella}
\author[inst2]{Qunyang Li}
\author[inst1,inst3]{Antonio Papangelo}
\affiliation[inst1]{organization={Politecnico di Bari, Department of Mechanics Mathematics and Management},
            addressline={Via Orabona 4}, 
            city={Bari},
            postcode={70125}, 
            country={Italy}}

\affiliation[inst2]{organization={AML, Department of Engineering Mechanics, Tsinghua University},
            city={Beijing},
            postcode={100084}, 
            country={China}}

\affiliation[inst3]{organization={Hamburg University of Technology, Department of Mechanical Engineering},
            addressline={Am Schwarzenberg-Campus 1}, 
            city={Hamburg},
            postcode={21073}, 
            country={Germany}}

\begin{abstract}
The contact between a rigid Hertzian indenter and an adhesive broad-band viscoelastic substrate is considered. The material behaviour is described by a modified power law model, which is characterized by only four
parameters, the glassy and rubbery elastic moduli, a characteristic exponent $n$ and a timescale $\tau_{0}$. The maximum adherence force that can be reached while unloading the rigid indenter from a relaxed viscoelastic half-space is studied by means of a numerical implementation based on the boundary element method, as a function of the unloading velocity, preload and by varying the broadness of the viscoelastic material spectrum. Through a comprehensive numerical analysis we have determined the minimum contact radius that is needed to achieve the maximum amplification of the pull-off force at a specified unloading rate and for different material exponents $n$. The numerical results are then compared with the prediction of Persson and Brener viscoelastic crack propagation theory, providing excellent agreement. However, comparison against experimental tests for a glass lens indenting a PDMS substrate show data can be fitted with the linear theory only up to an unloading rate of about  $100 \textrm{ $\mu$}$m/s showing the fracture process zone rate-dependent contribution to the energy enhancement is of the same order of the bulk dissipation contribution. Hence, the limitations of the current numerical and theoretical models for viscoelastic adhesion are discussed in light of the most recent literature results.
\end{abstract}

\begin{graphicalabstract}
\includegraphics[width=1\textwidth]{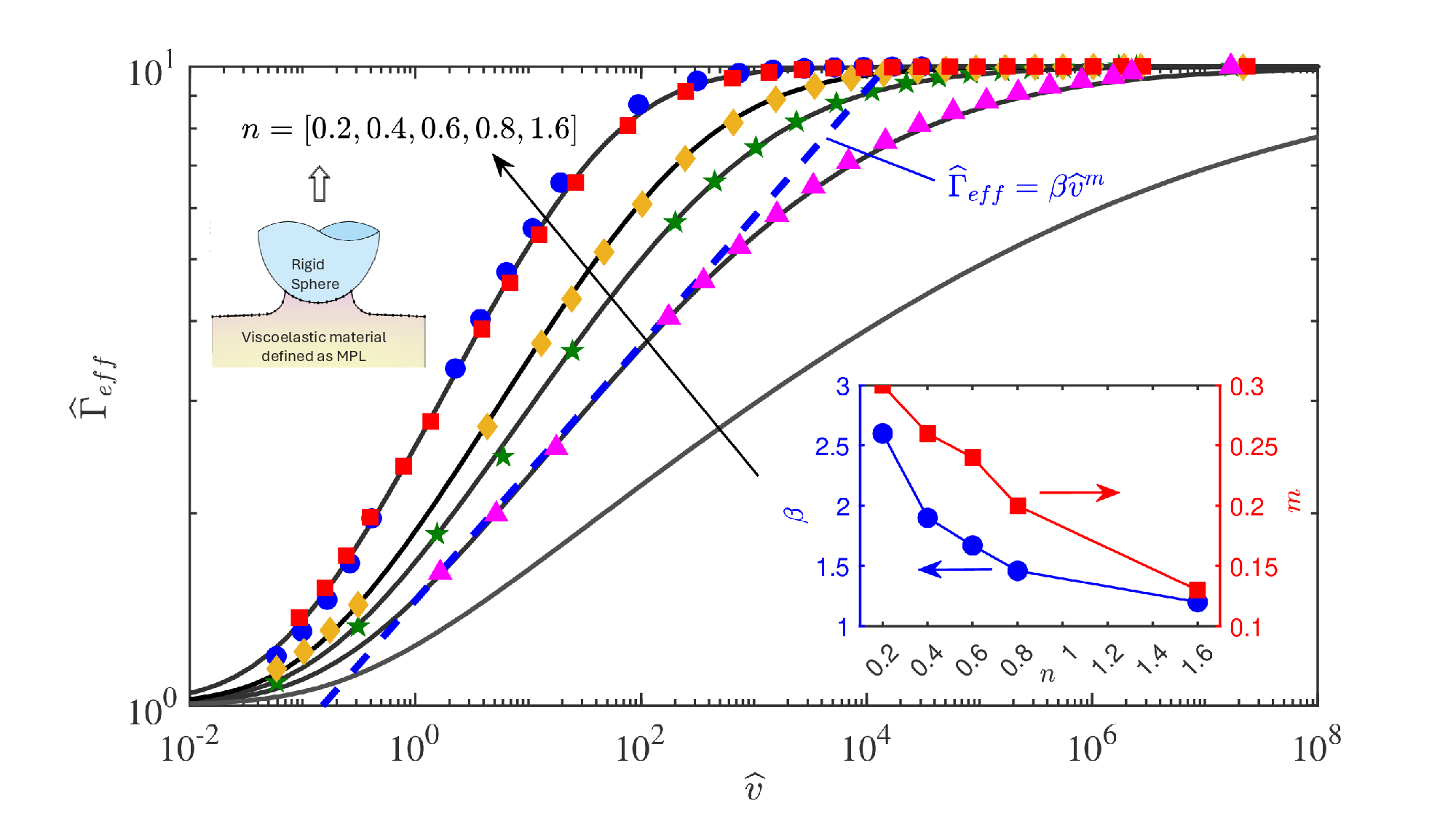}
\end{graphicalabstract}

\begin{highlights}
\item Numerical, theoretical and experimental investigations on viscoelastic adhesion enhancement
\item Maximum amplification is constrained by preload and unloading rate  
\item Closed form results for the effective surface energy in broad-band spectrum viscoelastic materials 
\item Bulk and fracture process zone contribution to the effective surface energy  
\item Experiments show limitations of classical linear theories
\end{highlights}


\end{frontmatter}


\section{Introduction} \label{sec:intro}
Understanding the adhesive behavior of soft materials such as polymers and
elastomers would be of interest in many engineering applications, ranging
from friction \citep{lorenz2015rubber, peng2021effect, nazari2024friction, mandriota2024adhesive},
gripping technologies \citep{shintake2018soft}, switchable adhesion \citep{kamperman2010functional, papangelo2017maugis,linghu2023overcoming, linghu2024fibrillar}, bio-mechanics \citep{felicetti2022tactile, forsbach2023two}, and
soft robotics \citep{mazzolai2019octopus, agnelli2021shape}. Since the fundamental
contribution of Johnson, Kendall, and Roberts, the JKR contact model \citep{johnson1971surface}, it is known that adhesion
of compliant elastic materials can be understood as a Griffith energy balance
between the elastic strain energy released as the crack advances and the
energy dissipated by the formation of new surfaces, as it is classically known
in Linear Elastic Fracture Mechanics \citep{maugis1992adhesion}. In this case, for quasi-static
conditions a thermodynamic work of adhesion $\Delta\gamma_{0}$, sometimes referred to as ``surface energy", can be defined that is independent of the rate at
which the remote load is applied.

If the crack is advancing in a viscoelastic media, care should be taken in
accounting for the dissipative effects introduced by viscoelasticity.
Currently, few theories attempt a description of this problem. The cohesive zone models \citep{schapery1975theory1,schapery1975theory2,greenwood1981mechanics,greenwood2004theory,schapery2022theory} attempt an accurate description of the adhesive interactions taking place at the crack mouth, more precisely within a "process zone" of length $l_{0}$ where the material bonds are actually disrupted and which is a function of the crack speed, resulting in the dependence on the speed of the effective surface energy $\Delta\gamma_{eff}$. A different model developed by \citet{ persson2005crack} (PB theory in the following) focuses on a steady-state moving crack. It establishes an energy balance between the power supplied by the applied load and the power dissipated by viscoelastic losses within the bulk material and the creation of new surfaces. By introducing a critical stress threshold for the rupture of material bonding, $\sigma_c$, it also introduces a typical lengthscale $l_0$ to calculate the dissipation in the bulk viscoelastic material \citep{ persson2005crack,persson2017crack, persson2021simple}.

Both cohesive and dissipation models really require a reference stress, or a reference length scale, which ultimately is used as a free parameter to fit the data points, and the linear theories result sometimes in nonphysical size for the fracture process zone at low speeds \citep{hui2022steady}, as we shall discuss with reference to our results in the Discussion paragraph. Schapery developed also more elaborate theories \citep{schapery1984correspondence} to include non linear stress-strain behaviour using $J$ integral and also far field viscoelasticity, which require obviously even more material characterization.

For a semi-infinite system, the cohesive-model approach and PB theory lead to a monotonic increase of the effective (or "apparent") surface energy $\Delta\gamma_{eff}$ with the velocity $v$ up to the theoretical ``high-frequency" limit of $\Delta\gamma_{eff}/\Delta\gamma_{0}=E_{\infty}/E_{0}$ being $E_{\infty}$ and $E_{0}$ respectively the glassy and the rubbery elastic moduli of the viscoelastic material \citep{ciavarella2021comparison}. For the unloading of a flat-punch from a viscoelastic substrate, both approaches have been recently revisited to include finite-size
effects, which have been showed to still give a monotonic increase of the
effective surface energy, but with a maximum amplification that is limited by
the system dimension providing $\max\left(  \Delta\gamma_{eff}/\Delta\gamma
_{0}\right)  <E_{\infty}/E_{0}$ \citep{maghami2024viscoelastic}. Even for systems that can be considered semi-infinite, the contact problem presents several
challenges as macroscopic adhesion is generally influenced by the indenter geometry \citep{papangelo2023detachment, maghami2024viscoelastic} and the contact history \citep{greenwood1981mechanics, violano2021jkr, violano2022size, afferrante2022effective, vandonselaar2023silicone}.

To precisely assess adhesion in soft polymers (silicone, rubber), the properties of the viscoelastic material need to be characterized. Several numerical and experimental works have tried to accurately {determine
the viscoelastic material response in the time domain \citep{wayne2011semi, lin2022extreme,dusane2023computational,qi2024mapping}, in the frequency domain \citep{huang2004measurements,efremov2017measuring} or using big data analysis and machine
learning algorithms \citep{saharuddin2020constitutive,hosseini2021optimized}.} All the
approaches reveal that real rubbers and elastomers have to be
characterized over a very wide range of frequencies which typically spans many orders of magnitude of the exciting frequency (broad-band behavior), and this, in turn, plays a crucial role in determining the bulk dissipation, hence the interfacial adherence force.

\begin{figure}[t]
\centering
\includegraphics[width=0.9\textwidth]{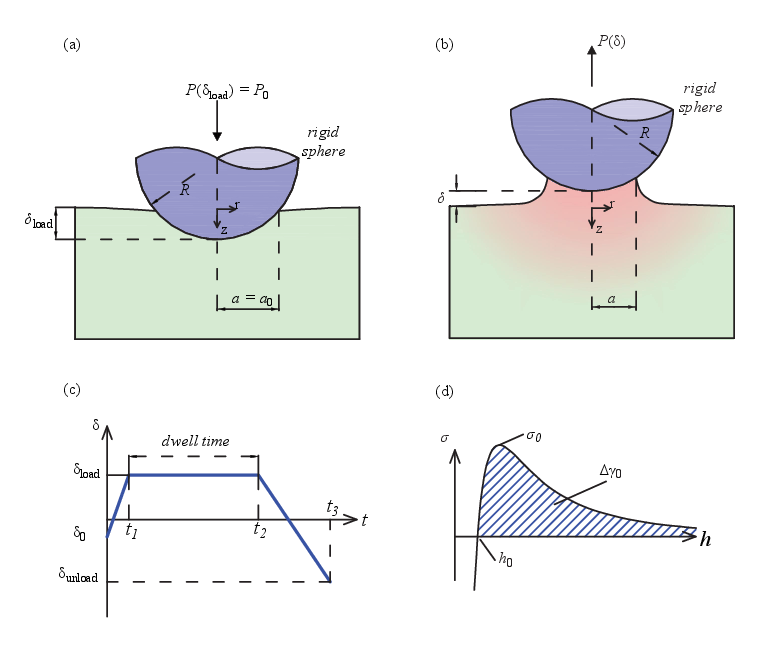}
\caption{Sketch of the geometrical model: (a) indentation phase, (b) unloading phase with a constant unloading rate; (c) loading protocol consisting of pre-loading, dwelling, and unloading; (d) the Lennard-Jones force-separation law used at the interface.}
\label{fig:identer}
\end{figure}

Here we shall consider the problem of a rigid sphere of radius $R$ that is unloaded from a fully relaxed broad-band viscoelastic adhesive half-space (see Fig.~\ref{fig:identer}) presenting and comparing numerical, analytical and experimental results. A few recent works \citep{violano2021jkr,violano2022adhesion,afferrante2022effective} have focused on the problem of the adhesion of a rigid Hertzian indenter unloaded from a viscoelastic substrate describing the material either by using (i) the classical three-element solid, also known as the Standard Linear Solid "SLS" (a spring in series with an element constituted by a dashpot and a spring in parallel), \citep{muser2022crack,violano2022adhesion,afferrante2022effective} or (ii) by considering the measured
response spectrum of the viscoelastic material used in the experimental campaign \citep{violano2021jkr, violano2022adhesion, vandonselaar2023silicone}. The limitation of the first approach is that the SLS has a narrow-band behavior, hence, although providing valuable
insights, it will be rarely useful for modeling the behavior of a real material. The limitation of the second approach is that the results obtained solve only the specific problem considered and its difficult to draw general conclusions.

The objectives of this work are: (i)\ to define a material model which may be
effectively and efficiently used for describing a real material viscoelastic
behavior with a minimal number of constants in both the time and frequency domain,
(ii) to determine, for the case of rigid Hertzian indenter-viscoelastic halfspace contact (see Fig. \ref{fig:identer}), how
the broad-band material behavior influences the maximum adherence force as
a function of the unloading rate (which is not the speed of the contact radius change, as we shall see) and of the preload, (iii)\ to provide closed
form results for the effective surface energy $\Delta\gamma_{eff}$ based on the
\citet{persson2005crack} theory which allows to faithfully reproduce the
numerical results together with their region of validity, (iv) to validate the proposed approach by comparing the numerical results with experimental data. 

The remainder of the paper is structured to address each objective outlined earlier: Section \ref{sec:MPL} provides a detailed description of the modified power law model used to characterize the viscoelastic material response; Section \ref{sec:numeric} presents, the Boundary Element Model developed for analyzing the adhesive contact problem along with extensive numerical results; Section \ref{sec:PB} introduces the developed analytical solution and demonstrates its applicability in characterizing adhesive contact problems; Section \ref{sec:experiment} focuses on the validation of the numerical results based on experimental outcomes; Section \ref{sec:disc} discusses the presented results in light of the recent Literature; the manuscript closes with the ``Conclusions", Section \ref{sec:conc}.

\section{Modified power law model} \label{sec:MPL}

The challenge behind the mechanical modeling of viscoelastic materials arises
because the mechanical response at time $t$ depends on the contact history, so
that the stress $\sigma\left(  t\right)  $ and strain $\varepsilon\left(
t\right)  $ should be determined by superposition:%
\begin{align}
\varepsilon\left(  t\right)   &  =\sigma\left(  0\right)  C\left(  t\right)
+\int_{0}^{t}C\left(  t-\tau\right)  \frac{d\sigma\left(  \tau\right)  }%
{d\tau}d\tau\label{creep-compliance}\;,\\
\sigma\left(  t\right)   &  =\varepsilon\left(  0\right)  R\left(  t\right)
+\int_{0}^{t}R\left(  t-\tau\right)  \frac{d\varepsilon\left(  \tau\right)
}{d\tau}d\tau\;,\label{relax}%
\end{align}
where $C\left(  t\right)  $ is the creep compliance function, giving the
strain response to a unit stress increment $\sigma\left(  t\right)  $ in
uniaxial loading conditions, and $R\left(  t\right)  $ is the relaxation
function giving the stress response to a unit strain increment $\varepsilon
\left(  t\right)  $ in uniaxial loading conditions. Alternatively,
viscoelastic materials can be characterized in the frequency domain. If a
sinusoidal stress $\sigma\left(  \omega\right)  $ at frequency $\omega$ is
applied to a viscoelastic specimen the resulting harmonic strain
$\varepsilon\left(  \omega\right)  $ will be delayed by a certain amount
$\delta$, hence the so-called complex modulus $\overline{E}\left(  \omega\right)
=\sigma\left(  \omega\right)  /\varepsilon\left(  \omega\right)  \ $can be
defined in the complex plane. Alternatively, in place of $\overline{E}\left(
\omega\right)  $ one may define its reciprocal, the $\overline{C}\left(
\omega\right)  =\varepsilon\left(  \omega\right)  /\sigma\left(
\omega\right)  $ which is the complex compliance. 

One approach to reproduce the broad-band response spectrum of a real
viscoelastic material \citep{vandonselaar2023silicone,lorenz2013adhesion} is to move from a SLS material model, which is constituted by a
spring in parallel with a single Maxwell element (a spring in series with a
dashpot), to the so-called Wiechert model constituted by a spring in parallel
with many Maxwell elements \citep{christensen2012theory}, so that several relaxation times can be included in the material representation. Very often the
number of elements needed for a faithful representation gets large enough so
that the model returns a very good representation of the material viscoelastic behaviour, but at the
same time it makes it difficult to extract general conclusions, due to the large number of fitting parameters determined. 

One option to overcome this difficulty is to rely on power law material models \citep{schapery1975theory1, persson2005crack, popov2010contact, bonfanti2020fractional, dusane2023computational}, which assume a certain power law function for the distribution of the
relaxation times. For example \citet{popov2010contact} proposes a model that is fully defined by 5 constants: the relaxed and glassy moduli, two characteristic times, and one exponent. \citet{schapery1975theory1} uses an approximation for the creep compliance function $C\left(  t\right)  =\left(
M_{e}+M_{1}t^{-p}\right)  ^{-1}$ which includes only three material constants $\left\{  M_{e},M_{1},p\right\}  $ and can describe well the behavior for very long times while being less accurate in describing the short-time material behavior. Furthermore, \citet{persson2005crack} consider a model where the retardation times are distributed as a power law in between two characteristic times and vanishes elsewhere.

In the following, we will consider and extend the Modified Power Law (MPL) material model
introduced by \citet{williams1964structural}, which is fully defined by a minimal set of four
parameters: the glassy $E_{\infty}$ and the rubbery $E_{0}$ moduli, a single
characteristic time $\tau_{0}$ and one exponent $n$. Closed-form results in both time and frequency domain that can be readily used for real viscoelastic material
characterization or as input in viscoelastic crack
propagation theories are provided in \ref{sec:appendixA}, while in the following the main results are reported.  

For the MPL material, the following relaxation spectrum $H(\tau)$ is assumed

\begin{equation}
H\left(  \tau\right)  =\left(  \frac{E_{\infty}-E_{0}}{\Gamma\left(  n\right)
}\right)  \left(  \frac{\tau_{0}}{\tau}\right)  ^{n}\exp\left(  -\frac
{\tau_{0}}{\tau}\right)\;,  \label{relspec}%
\end{equation}
where $\left\{  \tau_{0},n\right\}  $ are constants to be determined and $\Gamma(n)$ is the Gamma function. The
substitution of Eq. (\ref{relspec}) into Eq. (\ref{Ecmpl}) gives the complex modulus $\overline{E}\left(  \omega\right)  =E^{\prime
}\left(  \omega\right)  +\boldsymbol{i}E^{\prime\prime}\left(  \omega\right)
$:%
\begin{equation}
\overline{E}\left(  \omega\right)  =E_{0}+\left(  E_{\infty}-E_{0}\right)
\boldsymbol{i}\omega\tau_{0}\exp\left(  \boldsymbol{i}\omega\tau_{0}\right)
\mathbf{E}_{n}\left(  \boldsymbol{i}\omega\tau_{0}\right)\;,\label{Eomega}
\end{equation}
where $\mathbf{E}_{n}\left(x\right)  $ is the exponential
integral function of order $n>0$. \ref{sec:appendixA} provides closed-form results
for both the real $E^{\prime}\left(  \omega\right)  $ and the imaginary parts
$E^{\prime\prime}\left(  \omega\right)  $ of the complex modulus.

The relaxation function $R\left(  t\right)  $ is given by \citep{williams1964structural}: %

\begin{equation}
R\left(  t\right)  =E_{0}+\int_{0}^{\infty}\tau^{-1}H\left(  \tau\right)
\exp\left(  -t/\tau\right)  d\tau\;, \label{Rt2}%
\end{equation}
which, upon substitution of Eq. (\ref{relspec}) gives a very simple form:

\begin{equation}
R\left(  t\right)  =E_{0}+    \frac
{  E_{\infty}-E_{0}}{\left(1+t/\tau_{0}\right)^n}  \; \label{Rt3}%
\end{equation}
or in dimensionless form

\begin{equation}
\widehat{R}\left(  \widehat{t}\right)  =1 +    \frac
{  1/k-1}{\left(1+\widehat{t}\right)^n}  \;%
\end{equation}
being $\widehat{R}=R\left(  t\right)/E_0$, $\widehat{t}=t/\tau_0$ and $k=E_0/E_\infty$, which shows that at a given dimensionless time $\widehat{t}$ the material relaxation depends only on the parameters $\{n,k\}$.

Similarly, we can assume a modified power law distribution for the retardation spectrum

\begin{equation}
L\left(  \tau\right)  =\left(  \frac{C_{0}-C_{\infty}}{\Gamma\left(  n\right)
}\right)  \left(  \frac{\tau}{\tau_{0}}\right)  ^{n}\exp\left(  -\frac{\tau
}{\tau_{0}}\right)\;,  \label{retspec}%
\end{equation}
where $\left\{  \tau_{0},n\right\}  $ are constants to be determined. Hence
the complex compliance is (substitute Eq. (\ref{retspec}) into Eq.
(\ref{Dcmpl})):%

\begin{equation}
\overline{C}\left(  \omega\right)  =C_{\infty}+\frac{\left(  C_{0}-C_{\infty
}\right)  }{\boldsymbol{i}\omega\tau_{0}}\exp\left(  -\frac{\boldsymbol{i}%
}{\omega\tau_{0}}\right)  \mathbf{E}_{n}\left(  -\frac{\boldsymbol{i}}%
{\omega\tau_{0}}\right)\;,\label{Comega}
\end{equation}
where $\mathbf{E}_{n}\left( {x}\right)  $ is the exponential
integral function of order $n>0$, $C_{0}=1/E_{0}$ is the creep compliance in
the rubbery limit and $C_{\infty}=1/E_{\infty}$ is the creep compliance in
the glassy limit. Notice that once $\overline{C}\left(  \omega\right)  $ is
obtained, the complex modulus is also obtained as $\overline{E}\left(
\omega\right)  =1/\overline{C}\left(  \omega\right)  $ and vice-versa. The \ref{sec:appendixA} reports closed form results for both the real $C^{\prime}\left(
\omega\right)  $ and the imaginary part $C^{\prime\prime}\left(
\omega\right)  $ of $\overline{C}\left(  \omega\right)  $.

The creep compliance function $C\left(  t\right)  $ is given by (Williams, 1964):%

\begin{equation}
C\left(  t\right)  =C_{\infty}+\int_{0}^{\infty}\tau^{-1}L\left(  \tau\right)
\left(  1-\exp\left(  -t/\tau\right)  \right)  d\tau\;,\label{Ct2}%
\end{equation}
which, upon substitution of Eq. (\ref{retspec})\ into Eq. (\ref{Ct2})\ gives:%

\begin{equation}
C\left(  t\right)  =C_{0}-2\frac{\left(  C_{0}-C_{\infty}\right)  }%
{\Gamma\left(  n\right)  }\left(  \frac{t}{\tau_{0}}\right)  ^{n/2}%
\mathbf{K}_{n}\left(  2\sqrt{\frac{t}{\tau_{0}}}\right)\;,  \label{Ctgen}%
\end{equation}
where $\mathbf{K}_{n}\left(  x\right)  $ is the modified Bessel function of
the second kind. The dimensionless creep compliance function $\widehat{C}=C/C_0$ is

\begin{equation}
\widehat{C}\left(  \widehat{t}\right)  =1-2\frac{\left(  1-k\right)  }
{\Gamma\left(  n\right)  }  \widehat{t}^{n/2}
\mathbf{K}_{n}\left(  2\sqrt{\widehat{t}}\right)\;,  
\end{equation}

which shows that at a given dimensionless time $\widehat{t}$ the material creep depends only on the parameters $\{n,k\}$.

\begin{figure}[t]
\centering
\includegraphics[width=6in]{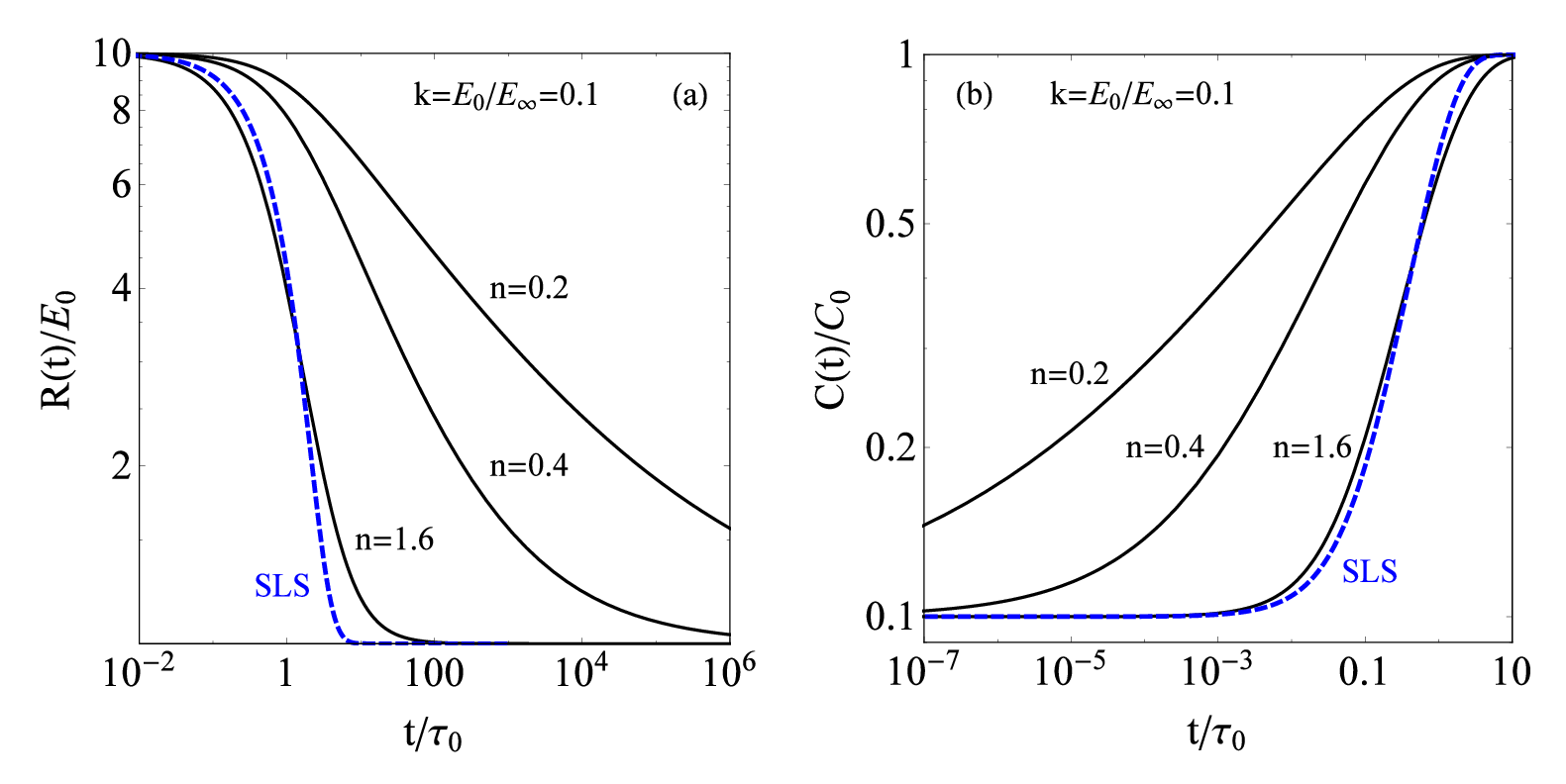}
\caption{Time evolution of (a) the relaxation function $R\left(  t\right)$ (Eq. 
\ref{Rt3}), (b) the creep compliance function $C(t)$ (Eq.
\ref{Ctgen}) for $n=[0.2,0.4,0.6]$ (solid black line) and a comparison with the behaviour of a SLS
viscoelastic material (dashed blue line).} \label{fig:MPLplot}
\end{figure}

In Section \ref{sec:experiment}, we will show that Polydimethylsiloxane (PDMS, 10:1 resin to curing agent ratio) one of the most common silicone-based polymer
used in soft contact mechanics \citep{shintake2018soft, sahli2019shear, oliver2023adhesion} has a characteristic exponent at room temperature of $n\simeq0.22$, which is
close to what \citet{williams1964structural} found for unfilled HC rubber. In Fig. \ref{fig:MPLplot}, we illustrate the time evolution of the relaxation and creep compliance functions of MPL for different exponents $n$ (solid black lines) alongside with a comparison of the SLS viscoelastic behaviour (blue dashed lines). Fig. \ref{fig:MPLplot} shows that in order to obtain a behavior close to a standard material, we should set $n\approx1.6$, which implies PDMS has a much broader spectrum with respect to the SLS. It is recalled that for a SLS the dimensionless creep compliance function is $\widehat{C}(\widehat{t})=[1-(k-1)\exp{(-\widehat{t})}]$.

\section{The numerical model} \label{sec:numeric}
{Let us consider the problem of a rigid sphere of radius $R$ that is unloaded from a fully relaxed}
viscoelastic adhesive half-space (see Fig.~\ref{fig:identer}). To model the adhesive contact
problem a numerical scheme based on the Boundary Element Method was
implemented in the software MATLAB, together with a time marching algorithm,
which follows the implementation by \citet{papangelo2020numerical, papangelo2023detachment}. In the numerical model it is assumed that the interaction between the sphere and the substrate is governed by a Lennard-Jones force-separation law \footnote{Strictly speaking Eq. (\ref{LJ}) would hold for infinite parallel planes, nevertheless in adhesive contact mechanics it is often assumed that Eq. (\ref{LJ}) holds also for slightly inclined surfaces, which is the so-called "Derjaguin approximation", see \citep{greenwood1997adhesion}.}:
\begin{equation}
\sigma\left(  h\right)  =-\frac{8\Delta\gamma_{0}}{3h_{0}}\left[  \left(
\frac{h_{0}}{h}\right)  ^{3}-\left(  \frac{h_{0}}{h}\right)  ^{9}\right]\;,
\label{LJ}%
\end{equation}
where $\sigma$ is the interfacial stress ($\sigma>0,$ when compressive), $h$
the local gap, $h_{0}$ the equilibrium distance with the surface energy $\Delta\gamma_{0}%
=\frac{9\sqrt{3}}{16}\sigma_{0}h_{0}$. The
gap function is then written as:
\begin{equation}
h(r,t)=-\delta+h_{0}+\frac{r^{2}}{2R}+u_{z}\left(  r,t\right)\;,  \label{h}%
\end{equation}
where $\delta>0$ when the sphere approaches the viscoelastic half-space, the sphere profile is approximated by a parabola, and
$u_{z}\left(  r,t\right)  $ is the deflection of the viscoelastic half-space,
which depends on the loading history (we have explicitly shown the dependence of $u_z$
on time $t$). The vertical deflections of the halfspace for an \textit{elastic} axisymmetric problem are
obtained as \citep{greenwood1997adhesion,feng2000contact}:%

\begin{equation}
u_{z}\left(  r\right)  =\frac{1}{E_{ps}}%
{\displaystyle\int}
\sigma\left(  s\right)  G\left(  r,s\right)  sds\;, \label{int}%
\end{equation}

where $G\left(  r,s\right)$ is the Kernel function:%

\begin{equation}
G\left(  r,s\right)  =\left\{
\begin{array}
[c]{cc}%
\frac{4}{\pi r}K\left(  \frac{s}{r}\right)  , & \qquad\qquad s<r\\
\frac{4}{\pi s}K\left(  \frac{r}{s}\right)  , & \qquad\qquad s>r
\end{array}
\right.  \label{kernel}%
\end{equation}
and $K\left(  k\right)$ is the complete elliptic integral of the first kind of
modulus $k$. Hence, according to the elastic-viscoelastic correspondence principle
in the form of Boltzmann integrals \citep{christensen2012theory}, the normal
displacements of the viscoelastic half-space $u_{z}(r,t)$ at time $t$, at
position $r$ will depend on the contact history as:
\begin{equation}
{u_{z}(r,t)=\int G\left(  r,s\right)  s\int_{-\infty
}^{t}C(t-\tau)\frac{d\sigma(s,\tau)}{d\tau}d\tau ds}\;. \label{int2}%
\end{equation}

where the moduli in the creep compliance function $C(t)$ should be consider in the plane strain conditions, i.e. $C^*_0=1/E^*_0=\frac{1-\nu^2}{E_0}$, being $\nu$ the Poisson ratio considered independent on the excitation frequency $\omega$. {The gap function Eq. (\ref{h}) is solved through a Newton-Raphson scheme on}
$N=M+1$ equally-spaced nodes, being $M$ the number of interfacial elements so that Eq.s (\ref{LJ},\ref{h},\ref{int2}) are satisfied. To
determine the half-space deflections Eq. (\ref{int2}) was discretized in time
and space. In time, we used a time marching algorithm with a time step $\Delta t$. In
space, we assumed the pressure distribution has a triangular shape over each
element, i.e. for the element $j$-th the pressure is $p_{j}$ at $r=r_{j}$ and
falls linearly to $0$ at $r=r_{j-1}$ and $r=r_{j+1}$, which is usually
referred as the ``method of the overlapping triangles" \citep{johnson1987contact}. Further
details of the numerical implementation can be found in Ref.s \citep{papangelo2020numerical,papangelo2023detachment}.

\begin{figure}
     \centering
     \includegraphics[width=0.7 \textwidth]{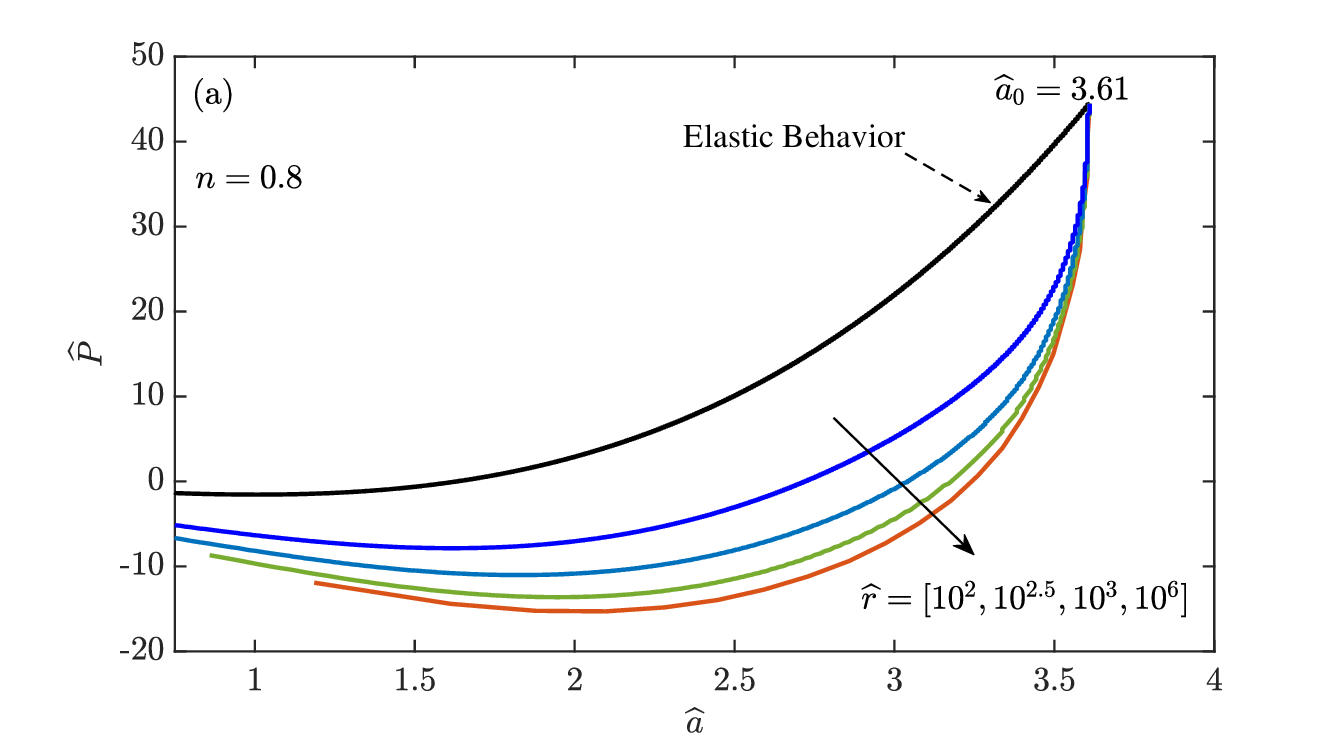}
          \includegraphics[width=0.72\textwidth]{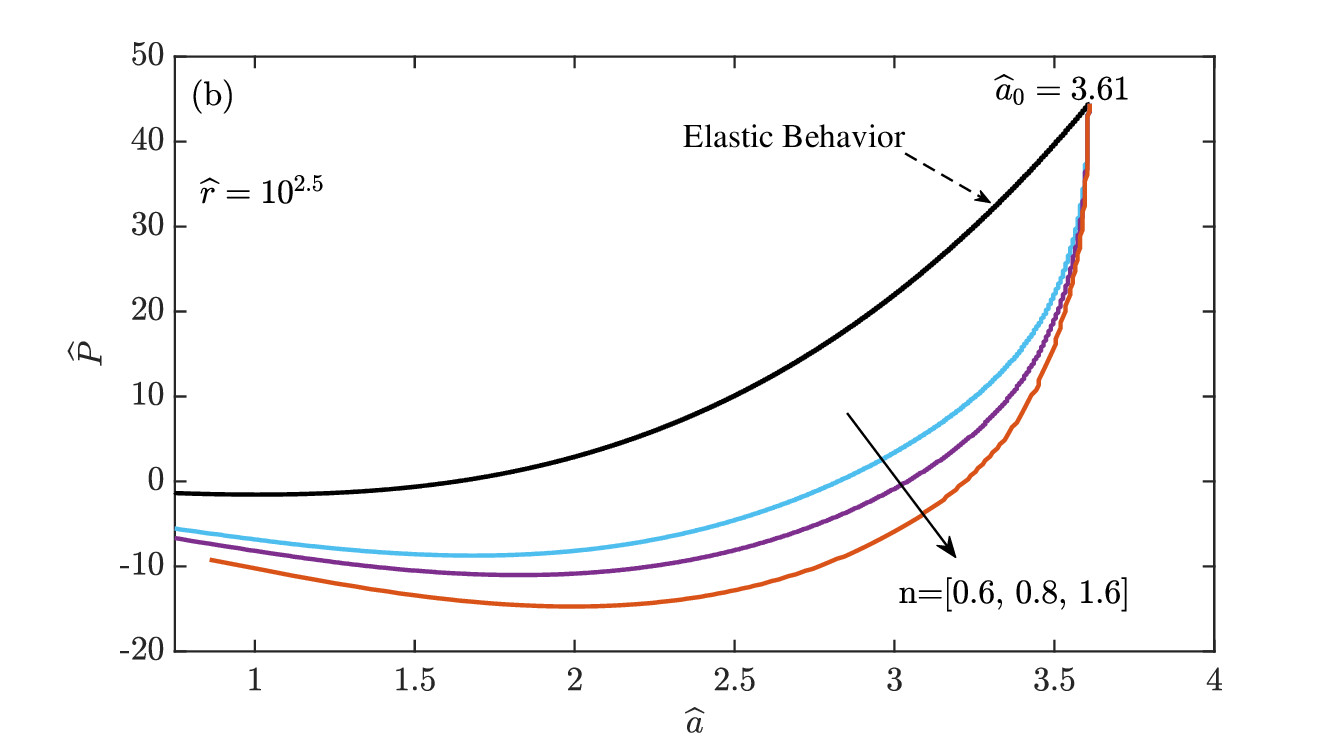}
    \includegraphics[width=0.7 \textwidth]{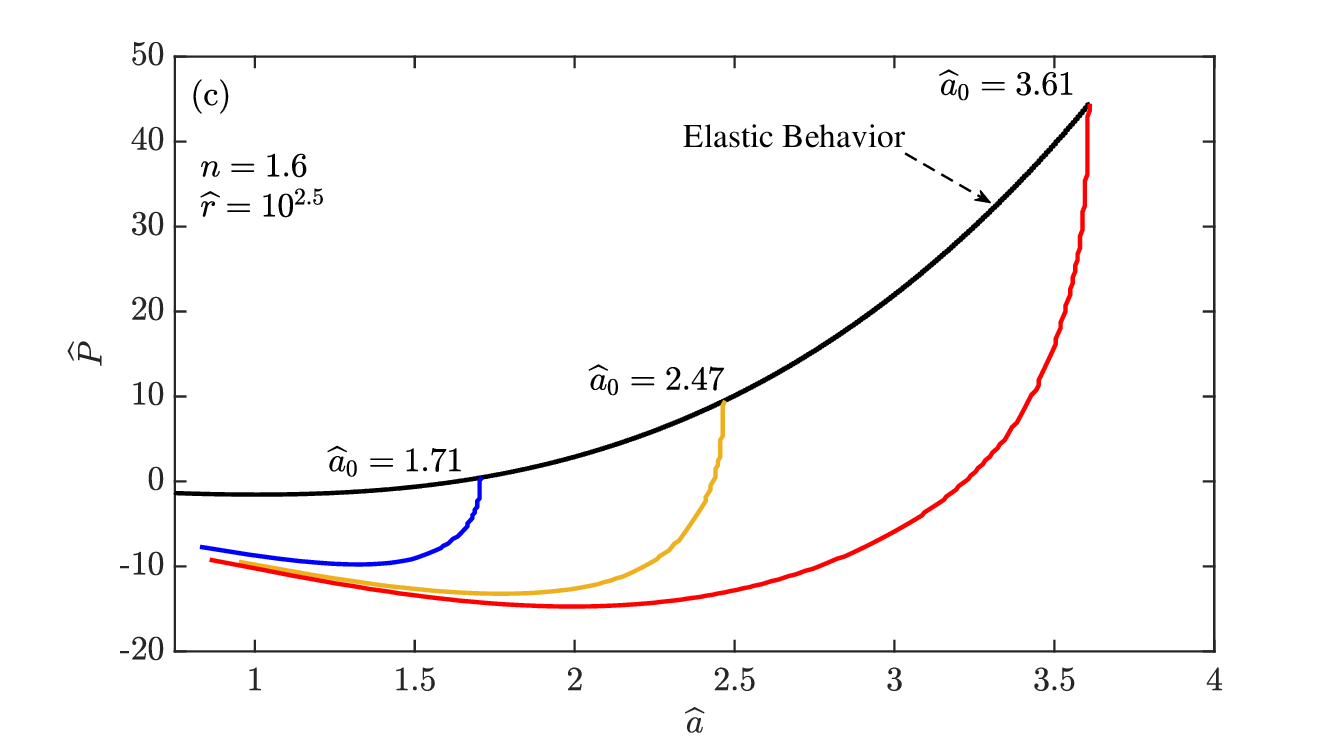}

              \caption{Dimensionless load $\widehat{P}$ versus the dimensionless contact radius $\widehat{a}$. (a) Material exponent $n=0.8$, initial contact radius $\widehat{a}_0=3.61$, unloading rates $\widehat{r}=[10^{2}, 10^{2.5}, 10^3, 10^6]$; (b) initial contact radius $\widehat{a}_0=3.61$, unloading rate of $\widehat{r} = 10^{2.5}$ for different material exponents $n=[0.6, 0.8, 1.6]$; (c) initial contact radii $\widehat{a}_0=[1.71, \;2.47,\;3.61]$, with a constant unloading rate of $\widehat{r}=10^{2.5}$ and material exponent $n=1.6$. For all the panels unloading starts from a fully relaxed substrate with $k=0.1$ and $\mu=3.24$:} \label{fig:3}

\end{figure}

\section{Numerical results}\label{sec:numres}

In the rest of the paper,
unless differently stated, the numerical results will be presented in
dimensionless notation, as follows:
\begin{equation}\label{eq:main}
\begin{split}
\widehat{\delta}=\frac{\delta}{(\pi^2 \Delta\gamma_0^2 R/{E_0^{*}}^2 )^{1/3}}\;;\quad  
\widehat{a}=\frac{a}{(\pi R^2 \Delta\gamma_0 / E_0^{*})^{1/3}}\;; \quad
\widehat{P}=\frac{P}{\pi \Delta\gamma_0 R}\;,
\end{split}
\end{equation}
being $\widehat{\delta}$ the dimensionless indentation, $\widehat{a}$ the dimensionless contact radius, $\widehat{P}$ the dimensionless normal load, $\widehat{P}_{po}=|\min\left(\widehat{P}\right)|$ the maximum detachment force, i.e. the pull-off force. Unless specified otherwise, our simulations employ $N = 500$ nodes. Let’s consider a sphere with an initial contact radius of ${a}_0$ is unloaded from a fully relaxed viscoelastic substrate, with various unloading rate ${r}$. This unloading process mimics experimental conditions: (i) indenting the viscoelastic substrate to a specified depth (${\delta}_{\textrm{load}}$), (ii) allowing dwell time for substrate relaxation, then (iii) unloading at a constant velocity ${r}$. The corresponding dimensionless unloading rate is defined as $\widehat{r}=r\tau/h_0$. Unless differently stated, the results provided in the following will refer to the Tabor parameter \citep{tabor1977surface} $\mu=\left(\frac{R{\Delta\gamma_0}^2}{{E^*_0}^2{h_0}^3}\right)^{1/3}=3.24$ and $k=E_0/E_{\infty}=0.1$.

\subsection{Dependence of the detachment force upon the loading protocol details}

{As it was discussed, the unloading rate $r$ has a crucial role in the mechanical response of viscoelastic materials. We examined our model for different unloading rates while Fig.~\ref{fig:3} (a) reports only four different unloading rates of $\widehat{r}=[10^{2}, 10^{2.5}, 10^3, 10^6]$. The sphere is unloaded from a fully relaxed substrate with the exponent material of $n=0.8$ and all the  unloading curves in Fig. \ref{fig:3} start from the initial contact radius of $\widehat{a}_0=3.61$. As anticipated in viscoelastic contact problems, the unloading rate significantly affects the unloading trajectory. Fast unloading boost viscoelastic dissipation at the crack tip which in turns gives a high pull-off load at detachment. Note that the elastic behavior observed in Fig. \ref{fig:3} corresponds to the initial state of the relaxed substrate.}

For the same unloading rate $\widehat{r}=10^{2.5}$, and starting from the same initial contact radius $\widehat{a}_0=3.61$ the unloading trajectory will be influenced by the response spectrum of the material. In particular, by using the MPL formulation for simulating a broad-band material (see Fig.~\ref{fig:MPLplot}), 
Fig.~\ref{fig:3} (b) shows the unloading curves for $n=[0.6, 0.8, 1.6]$, showing that for a given unloading rate the pull-off force generally increases by increasing $n$. As we will show later, this happens because the more narrow-band is the material response spectrum the faster the theoretical amplification $\Delta\gamma_{eff}/\Delta\gamma_0=E_\infty/E_0$ is reached as a function of the retraction rate.

\begin{figure} 
     \centering
        
        
        \includegraphics[width=6in]{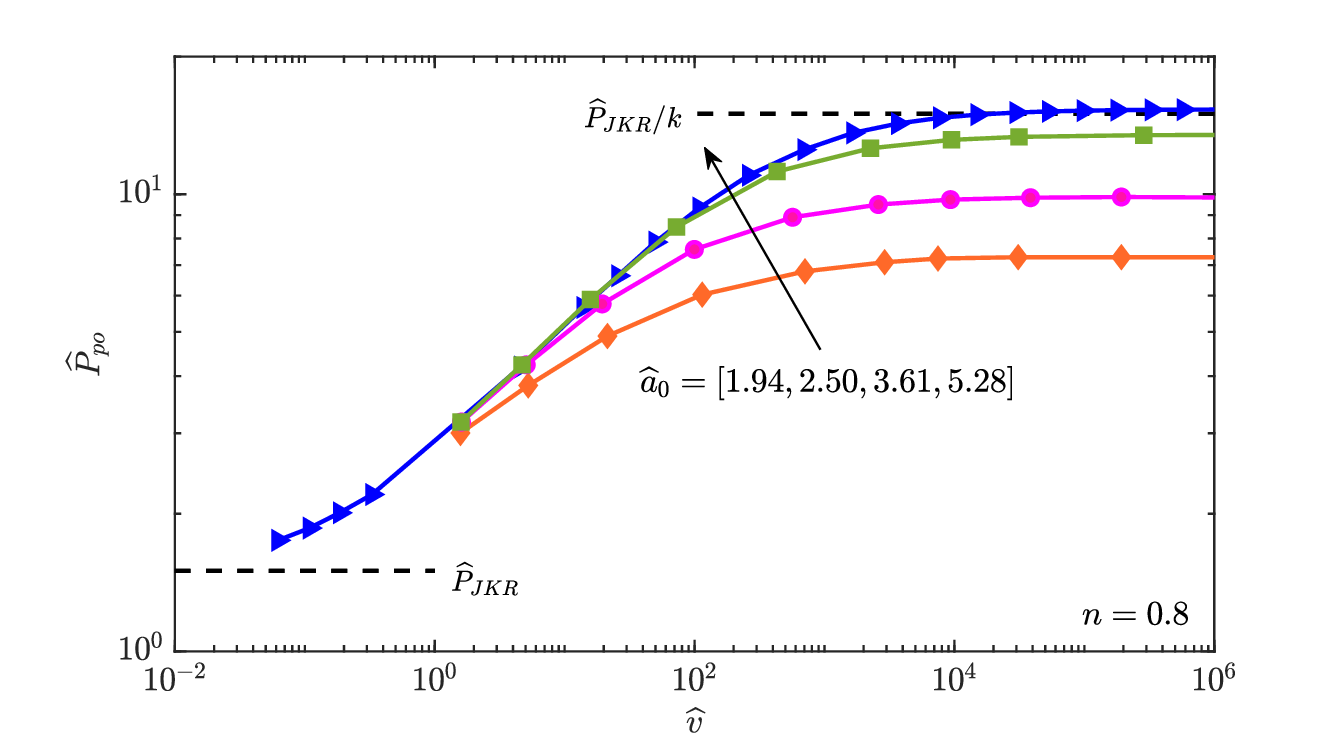}

              \caption{Normalized pull-off force as a function of the normalized crack velocity for a material with power law exponent $n=0.8$ and starting the unloading phase from a fully relaxed substrate with initial contact radii $\widehat{a}_0=[1.94,2.50,3.61,5.28]$.} \label{fig:aconv}
\end{figure}

Finally, the maximum adhesion force at detachment is significantly influenced by the the preload, as it was discussed in \citep{violano2022size,afferrante2022effective}. Changing the preload will affect the initial contact area, and also the initial indentation prior to unloading. To demonstrate this effect Fig.~\ref{fig:3} (c) shows the unloading trajectories of three different unloading curves from the same relaxed viscoelastic substrate (here $n=1.6$ and $\widehat{r}=10^{2.5}$) while using $\widehat{a}_0=[1.71, 2.47, 3.61]$ respectively. The larger is the initial contact radius the larger will be the pull-off force. Too small initial contact radius will give raise to finite size effects which will limit the possibility to enhance the pull-off force up to the theoretical limit predicted by viscoelastic crack propagation theories  \citep{persson2005crack,schapery1975theory1}. 

Hence, when looking for the maximum amplification of the pull-off force, care should be taken in selecting a large enough initial contact radius (or preload). Figure \ref{fig:aconv} shows the dimensionless pull-off force $\widehat{P}_{po}$ as a function of the dimensionless unloading rate $\widehat{r}=r\tau/h_0$ for a material with $n=0.8$ and by changing $\widehat{a}_0=[1.94,2.50,3.61,5.28]$. One realizes that while in the very beginning the enhancement curves all look similar, when the unloading rate starts to increase strong differences appear, with the curves referring to the smaller contact radius providing a much smaller enhancement, below the theoretical value even at the highest unloading rate tested. 

Although, the dependence of the pull-off force on the initial contact radius and on the unloading rate has been discussed for the paradigmatic SLS model \citep{violano2022size,afferrante2022effective}, it has remained unclear the importance of these parameters for real viscoelastic broad-band materials, hence we will address this question in the next subsection.

\subsection{Threshold contact radius}

\begin{figure}[t] 
     \centering
      
      \includegraphics[width=0.72\textwidth]{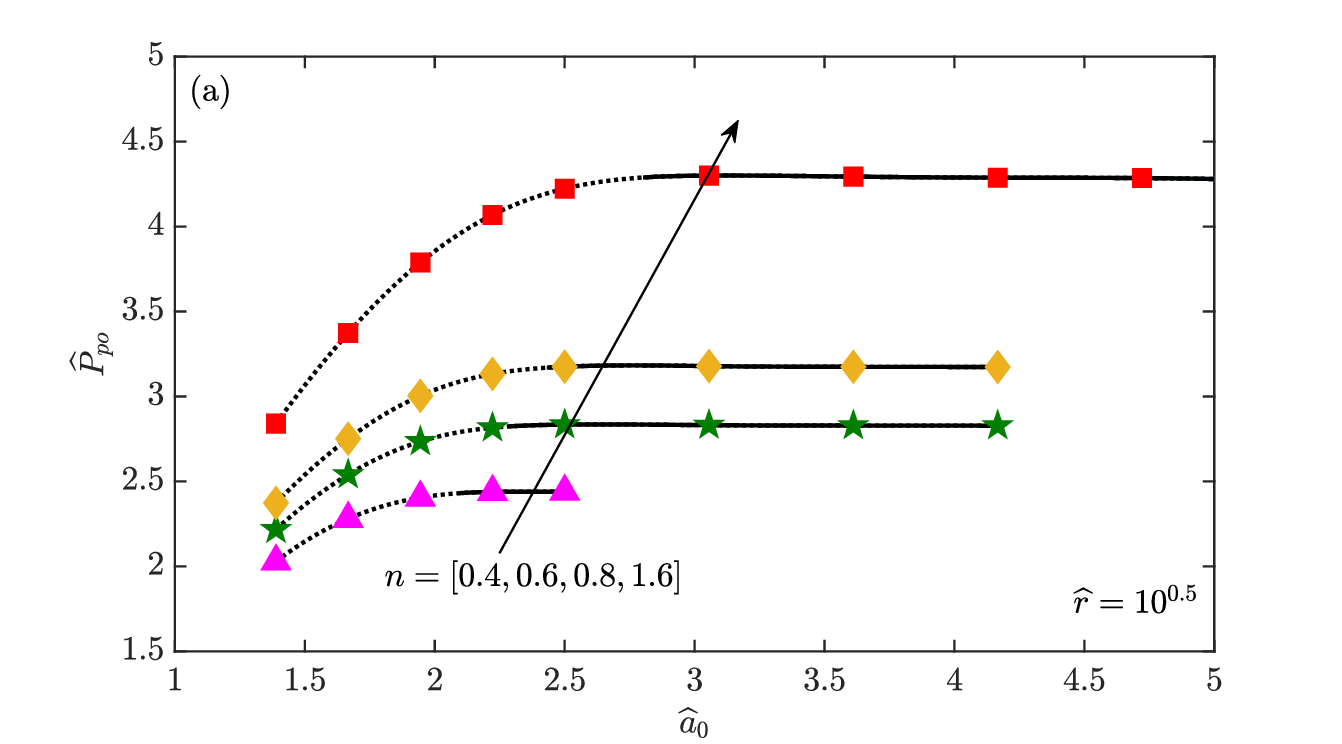}
      
      \includegraphics[width=0.72\textwidth]{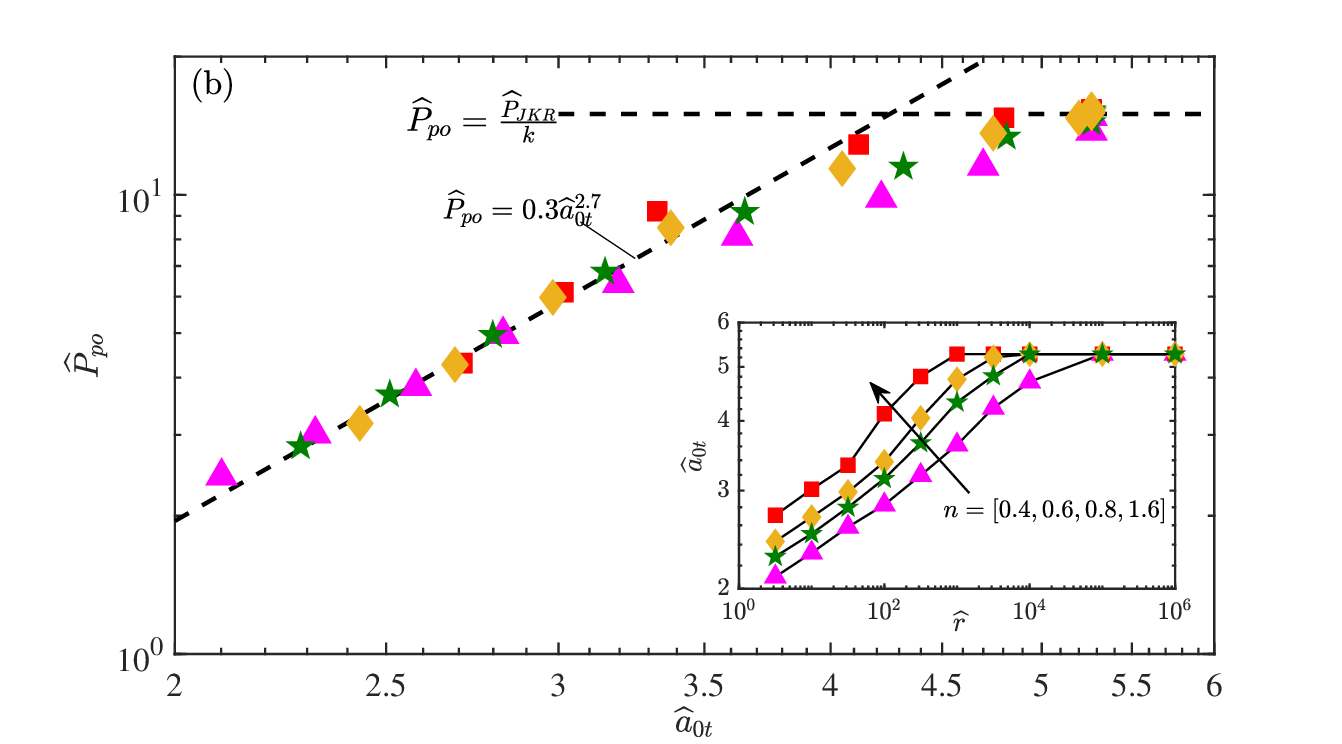}

              \caption{Threshold contact radius: (a) {Dimensionless pull-off force $\widehat{P}_{po}$ with respect to the normalized initial contact radius $\widehat{a}_0$ with the same unloading rate of $\widehat{r}=10^{0.5}$ for different material exponents $n=[0.4, 0.6, 0.8, 1.6]$, and $k = 0.1$}; (b, inset) Dimensionless threshold contact radius $\widehat{a}_{0t}$ with respect to the normalized unloading rate $\widehat{r}$ for different material exponents $n=[0.4, 0.6, 0.8, 1.6]$ and $k = 0.1$. (b, main figure) The same data reported in the inset are shown as dimensionless pull-off force $\widehat{P}_{po}$ with respect to the normalized threshold contact radius $\widehat{a}_ {0t}$. In all the panels triangle, star, diamond and square markers correspond respectively to $n=[0.4, 0.6, 0.8, 1.6]$.}\label{fig:Threshold}
\end{figure}

Here, the results of a comprehensive numerical campaign specifically designed for providing, at a given unloading rate, the minimal initial contact radius (or preload) needed to maximize adhesion in a rigid sphere/soft substrate contact are provided. Figure \ref{fig:Threshold} (a) shows the pull-off force obtained unloading the substrate at a given unloading rate $\widehat{r}=10^{0.5}$ as a function of the initial contact radius $\widehat{a}_0$ for $n=[0.4,0.6,0.8,1.6]$ (respectively triangle, star, diamond and square markers). The results show that as $\widehat{a}_0$ increases the pull-off force converges to a certain plateau and that in general, at a given $\widehat{r}$, broad-band materials (low $n$) will need a smaller initial contact radius to reach convergence of the pull-off force. Hence, in experiments, if the maximum adhesion is sought one must first perform a convergence study on the pre-loading conditions. In Fig. \ref{fig:Threshold} (a) we have used a spline to interpolate the simulated points (markers), then we have computed the derivative $d\widehat{P}_{po}/d\widehat{a}_{0}$ and set the condition $d\widehat{P}_{po}/d\widehat{a}_{0}<0.1$ to determine a  \textit{threshold contact radius} indicated by $\widehat{a}_{0t}$, above which we considered the pull-off force is converged. In Fig. \ref{fig:Threshold} (a) the black curves change from dotted to solid when the contact radius is grater than $\widehat{a}_{0t}$.

The results in Fig. \ref{fig:Threshold} (a) refer to a  particular unloading rate taken as a reference $\widehat{r}=10^{0.5}$. A convergence study was performed over about $5$ orders of magnitude in terms of unloading rate, as shown in the inset of Fig.~\ref{fig:Threshold} (b), where every marker shown corresponds to the threshold contact radius $\widehat{a}_{0t}$ obtained for that material exponent $n$ and at that given normalized unloading rate $\widehat{r} \in [10^0,10^6]$. The inset of Fig.~\ref{fig:Threshold} (b) explicitly shows a dependence on the viscoelastic material spectrum broadness (i.e. the exponent $n$), nevertheless if the data are represented as normalized pull-off force at convergence as a function of $\widehat{a}_{0t}$ they collapse for all the exponents $n$ into a single power law curve that we find to be $\widehat{P}_{po}=0.3\widehat{a}_{0t}^{2.7}$ (black dashed line in Fig.~\ref{fig:Threshold} (b)), which clearly saturates when the maximum enhancement $\widehat{P}_{po}=\widehat{P}_{JKR}/k=1.5/k$ is reached. Notice that, the smallest unloading rate considered in our analysis is $\widehat{r}\approx3$ (see Fig.~\ref{fig:Threshold} (b), inset) as for quasi static unloading the elastic solution is retrieved and $\widehat{P}_{po}$ will not depend on $\widehat{a}_{0}$.

Furthermore Fig.~\ref{fig:Threshold} (b) shows that the transition from the power-law behaviour to the plateau is faster for materials with large material exponent $n$ than for those characterized by low values of $n$, as a consequence of their narrow spectrum. Hence Fig.~\ref{fig:Threshold} (b) shows that regardless of the material model, the key parameter that determines the minimum contact radius $\widehat{a}_{0t}$ is the maximum amplification of the pull-off force that has to be reached. 

\section{Persson and Brener crack propagation theory for broad-band viscoelastic materials}\label{sec:PB}

In the previous sections, we have shown how the pull-off force depends
on the unloading rate and on the initial contact area for various exponent $n$ that characterize the broadness of the viscoelastic material response spectrum. Here, closed form solutions are obtained for the effective surface energy $\Delta\gamma_{eff}$ based on \citet{persson2005crack} crack propagation theory. It is useful to recall that for a Hertzian indenter, in the case of soft materials, the JKR model \citep{johnson1971surface} applies, which provides the pull-off force depends only on the sphere radius and surface energy $P_{JKR}=\frac{3}{2}\pi R \Delta\gamma$, hence, in the following, the normalized effective surface energy will be simply defined as $\widehat{\Gamma}_{eff}=\Delta\gamma_{eff}/\Delta\gamma_0 \simeq P_{po}/P_{JKR}$.

We note that our crack propagation formulation is the extension of \citet{persson2005crack} idea of equating the input power from the remote load to the power that is dissipated due to the generation of new surfaces and due to viscoelastic dissipation, so one can obtain the effective surface energy $\Delta\gamma_{eff}$ as \citep{persson2005crack}:
\begin{align}
\frac{\Delta\gamma_{eff}}{\Delta\gamma_{0}}  &  =\left[  1-\left(
1-\frac{E_{0}}{E_{\infty}}\right)  \int_{0}^{+\infty}\frac{L\left(
\tau\right)  }{\left(  1/E_{0}-1/E_{\infty}\right)  \tau}\left\{  \sqrt
{1+b^{-2}\left(  \tau\right)  }-b^{-1}\left(  \tau\right)  \right\}
d\tau\right]  ^{-1}\label{gammaeff}\;,\\
b\left(  \tau\right)   &  =\frac{2\pi v\tau}{l_{0}}\left(  \frac{\Delta
\gamma_{0}}{\Delta\gamma_{eff}}\right)\;.  \label{btau}%
\end{align}

\begin{figure} 
     \centering
       {\includegraphics[width=0.95\textwidth]{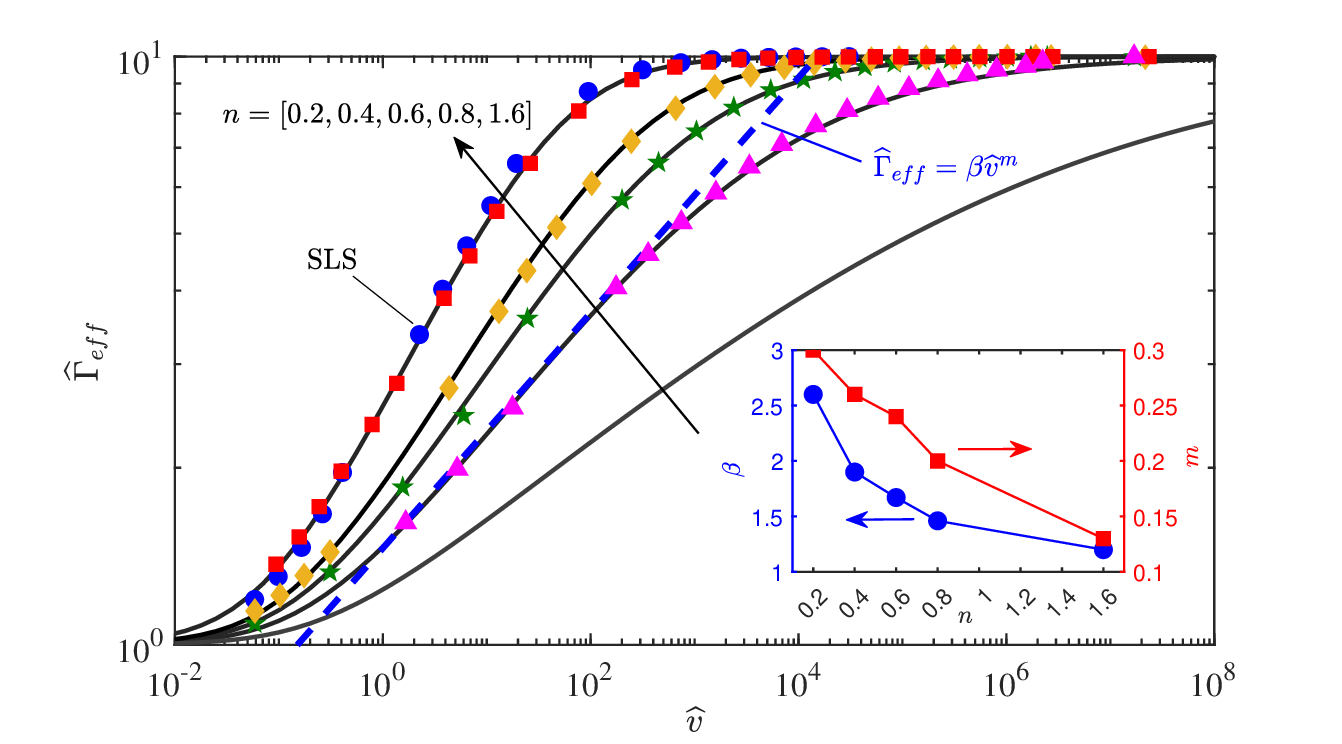}}
              \caption{Normalized effective surface energy $\widehat{\Gamma}_{eff}$ with respect to the normalized crack velocity $\widehat{v}$ for different power law material exponent $n=[0.4, 0.6, 0.8, 1.6]$, respectively triangle, star, diamond and square markers, and $k = 0.1$. The blue circle markers in the plot correspond to the SLS material. Solid lines stand for the PB model (Eq. \ref{PBmodel2}). The blue dashed line is a guide to the eye, showing the power law behaviour of the function $\widehat{\Gamma}_{eff}(\widehat{v})=\beta \widehat{v}^m$ in the intermediate velocity range. The inset depicts the fitting parameters $\{\beta,m\}$ for the values of $n$ tested.}\label{fig:PB_BEM}
\end{figure}

Introducing the dimensionless parameters:
\begin{equation}
\widehat{v}=\frac{v\tau_{0}}{l_{0}};\qquad \widehat{\tau}=\frac{\tau}{\tau_{0}}\;,%
\end{equation}\label{vc}
and substituting the retardation spectrum defined for the MPL material model in Eq. (\ref{retspec}) into Eq. (\ref{gammaeff}) one gets

\begin{equation}
\widehat{\Gamma}_{eff}=\left[  1-\left(  1-k\right)  \int_{0}^{+\infty}\frac{\widehat{\tau}^{n-1}%
\exp\left(  -\widehat{\tau}\right)  }{\Gamma\left(  n\right)  }\left[  \sqrt{1+\left(
\frac{\widehat{\Gamma}_{eff}}{2\pi \widehat{v}}\frac{1}{\widehat{\tau}}\right)  ^{2}}-\left(  \frac
{\widehat{\Gamma}_{eff}}{2\pi \widehat{v}}\frac{1}{\widehat{\tau}}\right)  \right]  d\widehat{\tau}\right]  ^{-1}\;,
\label{PBmodel}%
\end{equation}
which can be written as

\begin{equation}
\widehat{\Gamma}_{eff}=\left[  1-\left(  1-k\right)  I\left(  n,\widehat{v},\widehat{\Gamma}_{eff}\right)
\right]  ^{-1}\,, \label{PBmodel2}%
\end{equation}
where $I(n,\widehat{v},\widehat{\Gamma}_{eff})$ stands for the integral in Eq. \ref{PBmodel} which expression is given in closed form in the \ref{sec:AppendixB}. Following \citet{persson2005crack} original arguments, we determine the lengthscale $l_{0}$ equating the linear elastic fracture mechanics stress field to the critical stress $\sigma_{c}$ required to break the atomic bonds. Hence:

\begin{align}
\sigma_{c}  &  =\frac{K_{I}}{\sqrt{2\pi l_{0}}};\qquad K_{I}^{2}=\frac
{\Delta\gamma_{0}}{2E_{0}^{*}}\;,\\
l_{0}  &  =\frac{E_{0}^{^{*}}\Delta\gamma_{0}}{\pi\sigma_{c}^{2}}%
=\frac{E_{0}^{^{*}}\Delta\gamma_{0}}{\pi\left(  \alpha\sigma_{0}\right)
^{2}}\;,\label{eq:a0}
\end{align}
where $E_{0}^{^*}=\frac{E_{0}}{1-\nu^{2}}$ is the rubbery plain strain elastic
modulus of the halfspace, $K_I$ is the stress intensity factor in mode I and the ``$2$" in its expression takes into account that one of the contacting bodies
is rigid, while $\alpha$ {in Eq. (\ref{eq:a0})} is a coefficient of order unity to relate the critical stress $\sigma_{c}$ in PB theory to the $\sigma_{0}$ we are using in the numerical simulations that are based on the LJ force-separation law. Notice that for soft polymers ${l_{0}}/{h_{0}}\approx1$ hence $l_{0}$ should physically be of
the same order of the intermolecular distance.

Solving Eq. (\ref{PBmodel2}) for $n=[0.2, 0.4,0.6,0.8,1.6]$, $k=0.1$ and for varying crack velocity $\widehat{v}$ one easily find the results shown in Fig. \ref{fig:PB_BEM} (black solid lines). So, for a given effective energy, broad-band materials would require a much higher crack speed than for narrow-band materials. 
{The numerical results from the same set of parameters are shown in Fig. \ref{fig:PB_BEM} as markers ($n=[0.4,0.6,0.8,1.6]$, respectively triangle, star, diamond and square markers), where we find an excellent agreement with the analytical results by using $\alpha=\pi/9\simeq 0.3491$. It is reminded that the numerical results shown in Fig. \ref{fig:PB_BEM} have been obtained unloading a fully relaxed halfspace and are related to an initial contact radius exceeding the threshold value, i.e $\widehat{a}_0>\widehat{a}_{0t}$ (see Fig. \ref{fig:Threshold}). Numerical simulations conducted for $k=E_0/E_\infty=[0.01,0.05,0.1]$ confirmed that $\alpha\simeq\pi/9$ independently on the ratio rubbery to glassy modulus $k$. 

\begin{figure} 
     \centering
       \includegraphics[width=0.95\textwidth]{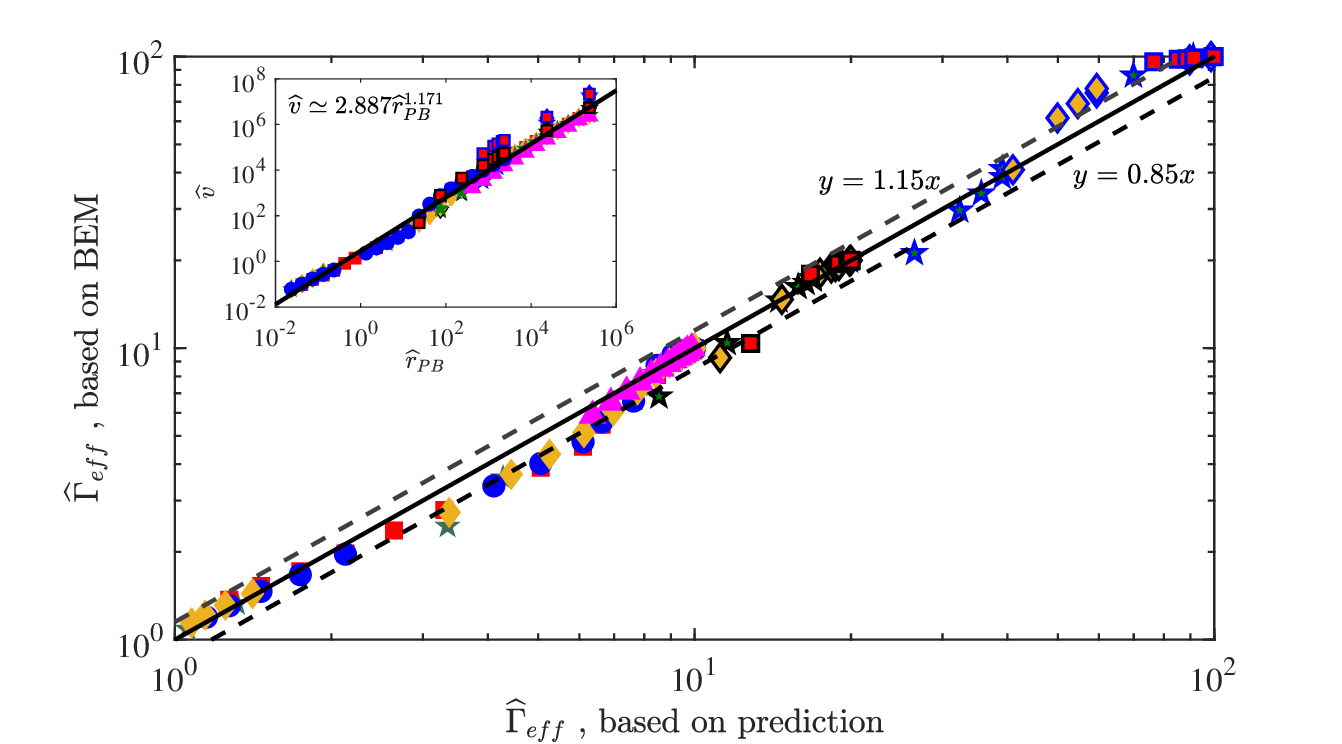}
              \caption{(main figure) Normalized effective surface energy $\widehat{\Gamma}_{eff}$ based on the numerical BEM simulations versus the normalized effective surface energy predicted by using the Eq.s (\ref{PBmodel2},\ref{eq:Vvsr}). (inset) Normalized crack velocity $\widehat{v}$ versus the normalized unloading rate ($\widehat{r}_{PB}$). In both panels the same simulations results are shown, in particular for different power law material exponent $n=[0.4, 0.6, 0.8, 1.6]$, respectively triangle, star, diamond and square markers, and $k=[0.01,0.05,0.1]$ respectively markers with a blue contour line, with a black contour line and without contour line. Blue circles stand for the SLS with $k=0.1$.} \label{fig:r_v}
\end{figure}

It is worth mentioning that the SLS is very often used as a paradigmatic model for a polymer viscoelastic behavior. As a comparison, Fig. \ref{fig:PB_BEM} reports the results obtained for a SLS as blue circles, which confirms the case of a SLS is close to $n=1.6$ and shows a notably large amplification of interfacial adhesion at relatively low crack speed if it is compared with broad-spectrum viscoelastic material. Our experimental results will show in Section \ref{sec:experiment} that 10:1 PDMS silicone have an exponent $n\simeq0.22$, which implies the maximum adhesion amplification may be observed only at unloading rates which are orders of magnitude larger than that needed for a SLS, which poses also questions about the practical feasibility of reaching so large retraction rates and possible nonlinear effects that may come into play, which will be discussed in the \textit{Discussion} section.
For a more convenient use of Eq. (\ref{PBmodel2}), the power law scaling of the effective surface energy in the intermediate velocity range is reported here as $\widehat{\Gamma}_{eff}=\beta \widehat{v}^m$ (see blue dashed line in Fig. \ref{fig:PB_BEM}), where the parameters ${\beta,m}$ can be found in Fig. \ref{fig:PB_BEM} inset.

The applicability of Eq. (\ref{PBmodel2}) for the prediction of the effective surface energy would remain limited by the fact that in all the viscoelastic crack propagation theories, including Eq. (\ref{PBmodel2}), the enhancement of the surface energy is a function of the crack velocity at pull-off which is generally not an input parameter in experiments and would be anyway difficult to control. Nevertheless, Fig. \ref{fig:r_v} shows in the inset that the crack velocity at pull-off $\widehat{v}$ scales approximately as

\begin{equation}
    \widehat{v}=2.887 \widehat{r}_{PB}^{1.171}\;, \label{eq:Vvsr}
\end{equation}
over about 10 orders of magnitude in term of unloading rate $\widehat{r}_{PB}$, where $\widehat{r}_{PB}=r\tau/l_0$. Figure \ref{fig:r_v} shows the numerical results obtained for the material exponents $n=[0.4,0.6,0.8,1.6]$, respectively triangles, stars, diamonds, squares (circles stand for the SLS material) and for $k=[0.01,0.05,0.1]$ respectively markers  with a blue contour line,  with a black contour line and without contour line. Filled blue circles stand for the SLS with $k=0.1$. Hence by using the Eq. (\ref{eq:Vvsr}) to estimate the crack speed at pull-off as a function of the retraction rate we have used Eq. (\ref{PBmodel2}) to predict the effective surface energy and compared with the numerical BEM results, which using the same symbols as in the inset, are shown in the main Fig. \ref{fig:r_v}. The solid black line represents the condition of perfect match between prediction and actual numerical results, while as a guide to the eye we have drawn also two dashed lines representing $\pm15\%$ error. Although the scaling may be improved by using more refined models, the use of Eq.s (\ref{PBmodel2},\ref{eq:Vvsr}) makes the estimate of the pull-off force straightforward based only on the material parameters and on the unloading rate. It is recalled that all the numerical results have been obtained for the Tabor parameter $\mu=3.24$, hence we expect Eq. (\ref{eq:Vvsr}) to be valid in the limit of short-range adhesion also referred to as the "JKR limit" \citep{johnson1971surface}.

\section{Experimental adhesion tests}\label{sec:experiment}
In the previous sections we have developed a general MPL material model capable of describing the viscoelastic behaviour of both narrow and broad band materials, then we have compared BEM numerical results with PB theory finding an excellent agreement. Finally, in this section the numerical predictions will be validated against experimental results.

A series of adhesion tests where performed using a smooth spherical lens loaded and unloaded from a soft viscoelastic substrate at various unloading velocities. The spherical lens was made of borosilicate crown glass (SLB-05-10P, Sigma Koki) with a nominal radius of R = 5.19 mm and the substrates were made of polydimethylsiloxane (PDMS, Sylgard 184, DowCorning Corporation)
with resin to curing agent weight ratio of 10:1. PDMS is a silicone elastomer well known to exhibit viscoelastic properties, as confirmed in several previous studies \citep{lorenz2013adhesion,vandonselaar2023silicone,violano2021jkr,petroli2022determination}. For the substrate material characterization a classical dog-bone shaped specimen was fabricated and used for dynamic mechanical analysis (DMA). All the samples were cured at 70 $^{\circ}$C for two hours on a heating table and then followed by natural cooling.

\subsection{Material characterization}

\begin{figure} 
    \centering
      \includegraphics[width=0.95\textwidth]{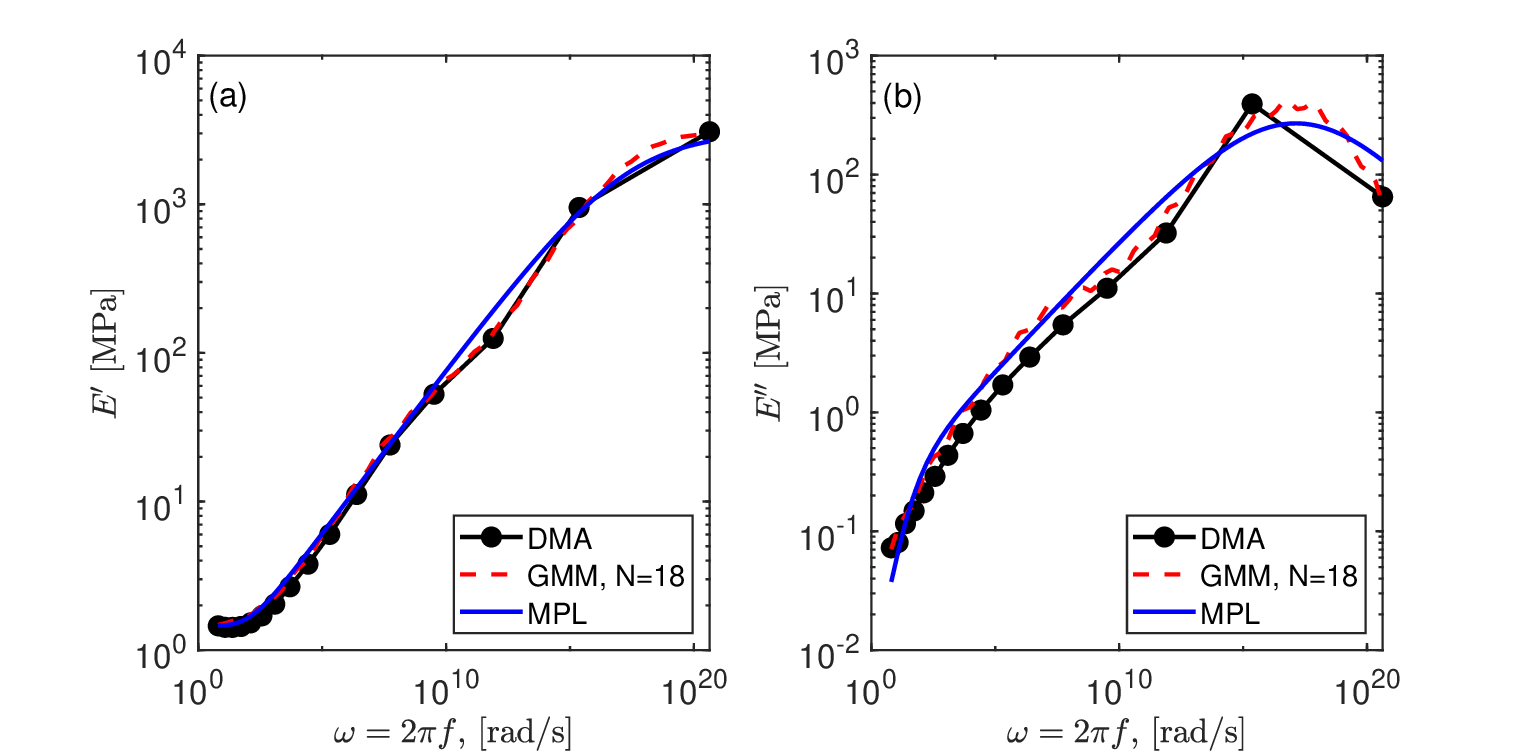}

\caption{(a) Real part of the complex elastic modulus
$E^{\prime}$ in the frequency domain at $T_{amb}=20\;{{}^\circ}$C. (b) Imaginary part $E^{\prime\prime}$ of the complex elastic modulus in the frequency domain at $T_{amb}=20$ ${{}^\circ}$C. In both panels: the black curve with circle markers stands for the experimental data, the blue curve for the fitted MPL material model and the red curve for the fitted GMM model with 18 arms (see Fig. \ref{fig:A1}).}\label{fig:EE}
\end{figure}

{The DMA test was performed using a DMA850 (TA Instruments) to characterize the viscoelastic properties of the PDMS.} The dog-bone-shaped specimen had cross-sectional dimensions of $3.86$ mm in width and $0.75$ mm in thickness. Temperature sweeps were conducted at a fixed frequency of $f =1$ Hz and a strain amplitude of $\epsilon = 0.1 \%$. The temperature runs from $-130\;{{}^\circ}$C to $20\;{{}^\circ}$C with $10\;{{}^\circ}$C step size. {To move} from temperature to frequency domain we used the WLF time-temperature superposition \citep{WLF}, hence the shift factor is defined as
\begin{equation}
\log_{10}a_{T}=\log_{10}\frac{f_{T_{g}}}{f_{T}}=\frac{-17.44\left(
T-T_{g}\right)  }{51.6+T-T_{g}}\,, \label{aT}%
\end{equation}
where $f_{T}$ is the frequency at the temperature $T$ and $T_{g}$ is the glass transition temperature. For the PDMS substrate we assumed $T_{g}=-115^\circ$C, which agrees well with the results reported in Ref. \citep{vandonselaar2023silicone} for the same material. Furthermore, we note that using $T_{g}=-115^\circ$C our measurements of the complex modulus $\overline{E}$ also satisfy the Kramers-Kronig (KK) relation \citep{pritz2005unbounded}

\begin{equation}
E^{\prime\prime}\left(  \omega\right)  =-\frac{2\omega}{\pi}\int_{0}^{+\infty
}\frac{E^{\prime}\left(  u\right)  }{\omega^{2}-u^{2}}du\,,
\end{equation}
where $\omega=2\pi f$ is the angular frequency and the integral should be intended as its Principal Value \citep{pritz2005unbounded}.%

The experimental data for the complex modulus were shifted to $T_{amb}=20\;^\circ$C by using Eq. (\ref{aT}) and fitted using Eq. (\ref{Comega}), which is written in terms of the complex compliance $C(\omega)$ as that is the function needed in the numerical BEM implementation (see Eq. (\ref{int2})). Figure \ref{fig:EE}, panels (a)-(b), shows the complex modulus $\overline{E}\left(  \omega\right)  =1/\overline{C}\left(  \omega\right)  =E^{\prime}\left(
\omega\right)  +\boldsymbol{i}E^{\prime\prime}\left(  \omega\right)$ as obtained experimentally (black solid curve with circle markers) and as fitted by the MPL material model (blue solid curve). For PDMS we found

\begin{equation}%
\begin{array}
[c]{c}%
E_{0}=1.458 \text{ MPa}\\
E_{\infty}=3.089\ast10^{3}\text{ MPa}\\
n=0.2207\\
\tau_{0}=0.01876\text{ s}%
\end{array}
\end{equation}

For comparison purpose, the result that would have been obtained by fitting the experimental data using a Generalized Maxwell Model (GMM, also known as the "Wiechert model") with 18 arms, hence 37 constants, is also shown in Fig. \ref{fig:EE} as a dashed red curve. One realizes that both the GMM and the MPL models give a fair representation of the material behavior, although the MPL model is simpler to use, and the four parameters used in the fitting procedure $\left\{E_{0},E_{\infty},n,\tau_{0}\right\}$ have a straightforward physical interpretation.

\subsection{Experimental setup and comparison}
A custom-designed adhesion test instrument, based on the tribometer platform (NTR2, CSM Instruments), was constructed to measure the pull-off force. As illustrated in Fig. \ref{fig:setup} (a), the lens was rigidly fixed to the force sensor. The PDMS substrate was positioned above a transparent rigid box, with the contact interface observable through a camera via a prism mounted inside the box. The pull-off tests comprised three sequential steps: loading, dwelling, and unloading. Initially, the lens was gradually loaded against the PDMS substrate with a preload force denoted as $P$, followed by a dwell period of 60 seconds to ensure complete relaxation of adhesive contact. Subsequently, the lens was pulled out at a fixed unloading rate, $r$. Throughout the entire process, the normal force was recorded and the pull-off force represents the absolute minimum normal force.
Firstly, we measured the variation of the normal force with the contact radius, $a$, at a very low unloading rate $r=0.98 \mu$m/s  to determine the interfacial parameters, as shown in Fig. \ref{fig:setup} (b). By fitting the relationship between the normal force and contact radius {using} Carpick’s method \citep{carpick1999general}, we estimated the intrinsic work of adhesion ${\Delta\gamma}=0.152\; \text{J/m}^2$ and the Tabor parameter $\mu=2.05$. Next, we conducted tests by varying the unloading rate $r$. The lens is brought into contact with the PDMS substrate and loaded to the preset preload $P_0=1.5$ mN. After a 60-second dwell period, the lens is moved upward until {the} contact is broken and the lens is pulled off from the substrate.
\begin{figure} 
     \centering
     
     \subfloat[]{\includegraphics[width=0.48\textwidth]{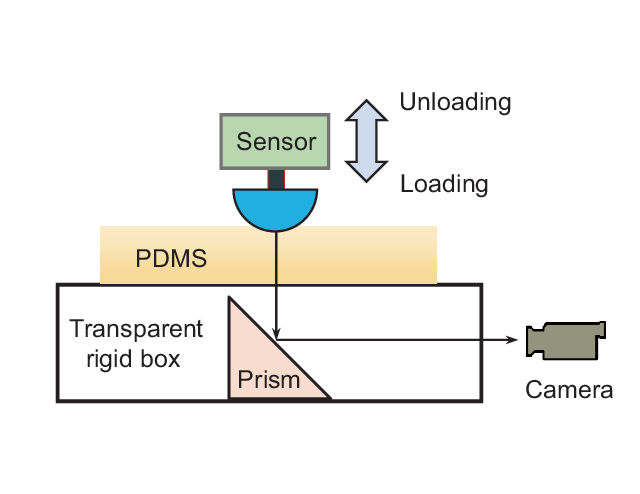}}
     \subfloat[]{\includegraphics[width=0.48\textwidth]{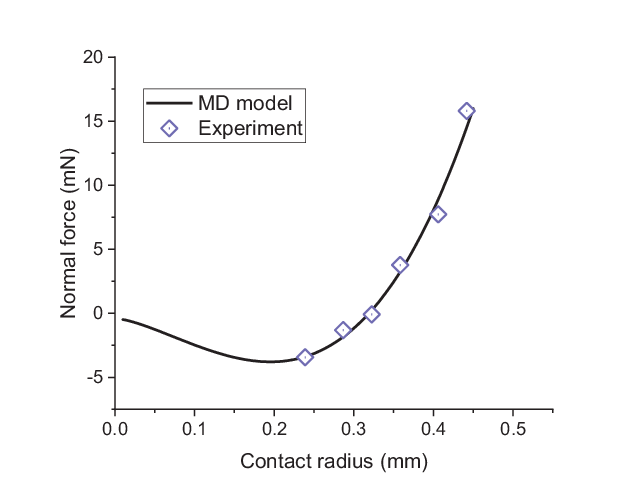}}
              \caption{ (a) Schematic of experimental setup for adhesion tests; (b) Variations of normal force with contact radius $a$ at a very low unloading rate to determine the interfacial parameters.} \label{fig:setup}
\end{figure}
We used our numerical BEM code, using the MPL material model for the viscoelastic substrate, to predict the pull-off force during the unloading process. The comparison with experiments leads to the result shown in Fig. \ref{fig:comparison}, where the pull-off force, ${P}_{po}$ [mN], is plotted as a function of the unloading rate, $r$ [$\mu$m/s] (the red squares stand for the experimental data, the black solid line for the numerical results). According to Fig. \ref{fig:comparison}, the experimental results confirm an increase in the pull-off force with increasing unloading rates. While there is good agreement between numerical and experimental results in the range of retraction rates $r=[1,100]$ $\mu$m/s, the experimental data exhibit marked higher values than the numerical predictions for high values of the unloading rate $r>100$ $\mu$m/s, which agrees well with other published experimental results \citep{vandonselaar2023silicone,tiwari2017effect}. For having a good fit of the low speed experimental results, we set $h_0=30.8$ nm, which is discussed in detail in the Discussion section. 

\begin{figure} 
     \centering
     {\includegraphics[width=0.75\textwidth]{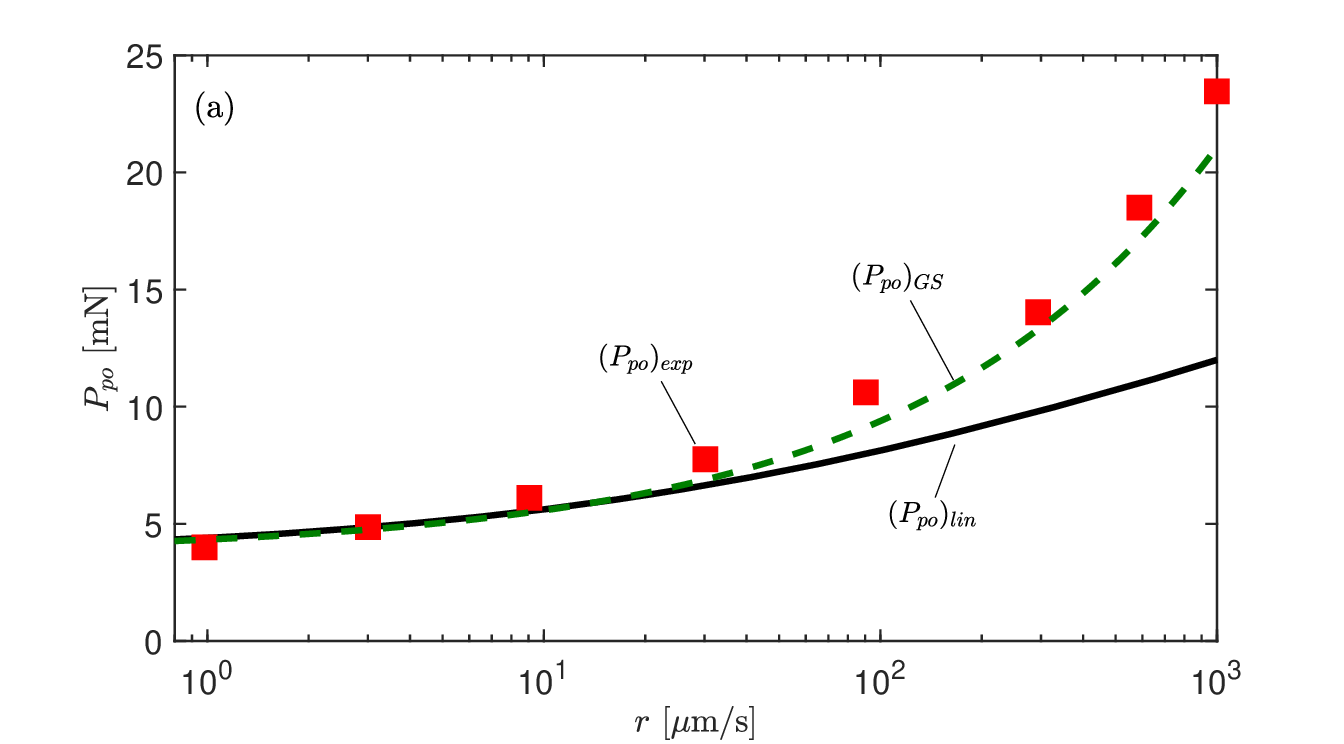}}
     {\includegraphics[width=0.75\textwidth]{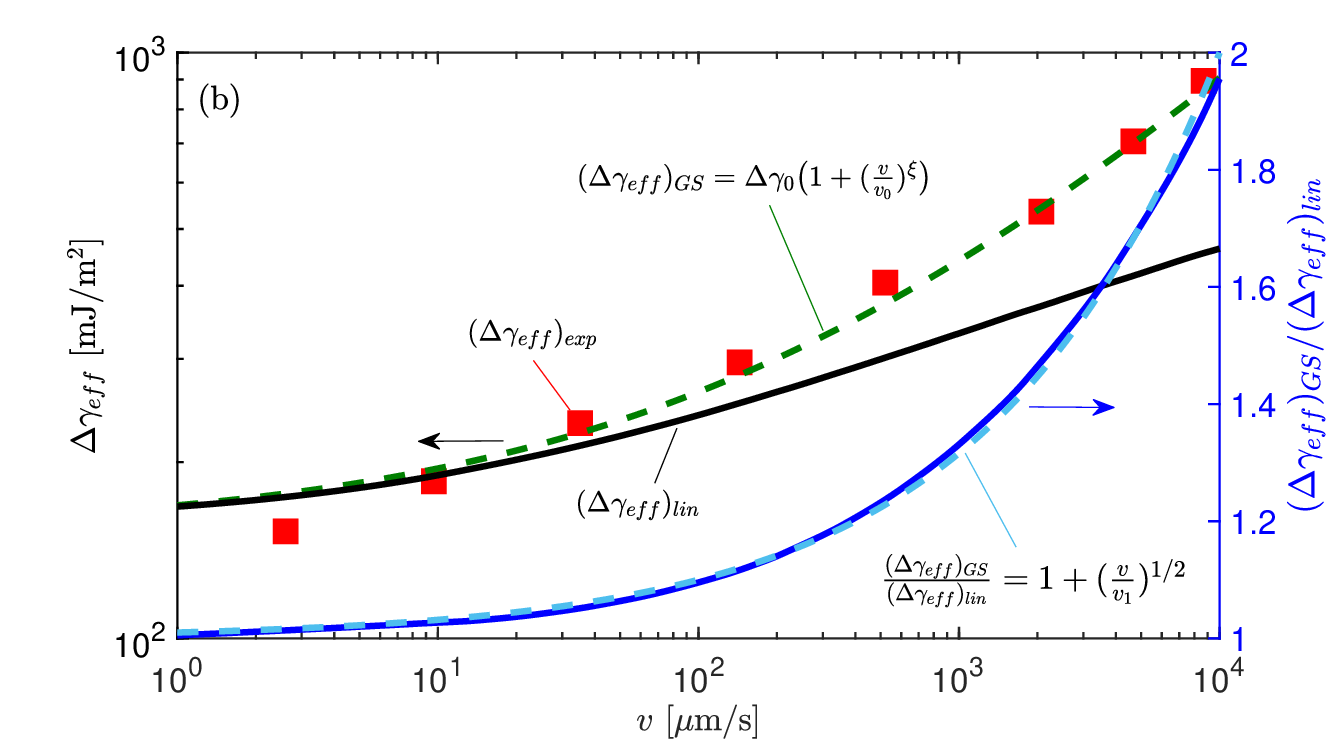}}
              \caption{(a) Pull-off force as a function of the unloading rate: comparison between numerical (from the BEM code, solid black line, labelled $(P_{po})_{lin}$) and experimental (red square markers) results for preload $P_0=1.5$ mN, $R=5.19$ m$^{-3}$, $\Delta\gamma_0=152.3$ mJ/m$^2$, $h_0=30.8$ nm. The green dashed line was obtained using the effective surface energy fitted on the experimental results  labelled as $(\Delta\gamma_{eff})_{GS}$ and shown in panel (b). (b, left y-axis) Effective surface energy from: the experimental results (red squares, labelled $(\Delta\gamma_{eff})_{exp}$), the fit of the experimental data using a Gent and Schulz power law model (Eq. (\ref{GS}), where $v_0=213.5$ $\mu$m/s and $\xi=0.4154$, dashed green line, labelled $(\Delta\gamma_{eff})_{GS}$), the prediction obtained for the PDMS substrate using the \textit{linear} numerical BEM model (solid black line, labelled $(\Delta\gamma_{eff})_{lin}$). In this respect we used our approximate Eq. (\ref{eq:Vvsr}) to determine the crack velocity at pull-off starting from the experimental retraction rate. (b, right y-axis) The ratio between $(\Delta\gamma_{eff})_{GS}/(\Delta\gamma_{eff})_{lin}$ for which a power law fit is provided $\frac{ (\Delta\gamma_{eff})_{GS} } { (\Delta\gamma_{eff})_{lin} }=1+(\frac{v}{v_{1}})^{1/2}$ where $v_1=10^4 \mu$m/s (dashed pale blue curve).}  \label{fig:comparison}
\end{figure}

\section{Discussion}\label{sec:disc}

\subsection{On the possibility to reach the maximum adhesion enhancement}

According to viscoelastic crack propagation theories \citep{persson2005crack,schapery1975theory1,schapery1975theory2} the maximum enhancement of the pull-off force is $\widehat{P}_{po}=\widehat{P}_{JKR}/k$, hence based on the results reported in Section \ref{sec:numres}, one can estimate that for a PDMS material with $k\simeq4.73*10^{-4}$ the maximum amplification of the pull-off force will be reached for $\widehat{a}_{0t}\simeq30.9$. Using the interfacial and material properties we have found for PDMS $(\Delta\gamma_0=152.3 \textrm{ mJ/m}^2, \nu=0.5, E^*_0\simeq1.94 \textrm{ MPa})$ and for $R=5.19 \textrm{ mm}$ gives an initial contact radius of ${a}_{0t}=5.8$ mm, which is larger then the sphere radius and even considering a parabolic (Hertzian) profile certainly outside the limit of validity of the hypothesis of small deformations, which raises doubts about the practical feasibility of reaching the maximum amplification factor predicted by crack propagation theories.  

Another consideration to be made is related to the unloading rate that would be needed to reach the maximum adhesion amplification. By using the results reported in Section \ref{sec:experiment}, one estimates that to reach the maximum amplification for a PDMS substrate one would need to unload the substrate at $\widehat{v}\approx10^9$ (see Fig. \ref{fig:PB_BEM}), which using $\{ l_0=30.0 \textrm{ nm}, \tau_0=0.01876 \textrm{ s} \}$ together with Eq. (\ref{eq:Vvsr}) gives the dimensional unloading rate of about $r\approx31.4$ m/s, which is about 4 orders of magnitude larger than the maximum unloading velocity usually used in adhesion experiments \citep{tiwari2017effect,vandonselaar2023silicone}, provided also the limitations introduced by the inertia of the motorized linear stages. Hence, insufficient preload and unloading rates used in experiments may partially explain why, in Literature, measurements of very large enhancement factors, even close to $1/k$, are missing (see for example \citet{vandonselaar2023silicone,tiwari2017effect}).

On the other hand, our numerical and experimental results seem to be in agreement with the  experimental adhesion tests reported in Refs. \citep{vandonselaar2023silicone,tiwari2017effect} for a similar PDMS material, where they also found that PB theory agreed well with experimental observations only up to about $r\approx100 \textrm{ $\mu$}m/s$. This may suggest that both numerical and theoretical models are lacking of essential phenomena to describe the detachment process at high retraction rates. At present, a few hypotheses have been formulated, ranging from the possibility of nonlinear dissipative phenomena, happening within the process zone, like (i) cavitation and stringing, (ii) extraction of non-cross linked polymeric chains from the substrate, (iii) temperature dependence of the material behaviour at the crack tip, (iv) the nonlinear behaviour of the material at the large strains ($\approx10\%$) experienced close to the crack tip \citet{vandonselaar2023silicone,tiwari2017effect}, of course not included into the (linear) theoretical and numerical models, which is discussed in the next subsection.


\subsection{Energy dissipation within the process zone}

To obtain a satisfactory fit of the experimental data at low unloading rates we set $h_0=30.8$ nm. Notice that from quasi-static experiments, using the definition of the Tabor parameter we would have obtained a much larger equilibrium distance $h_0=1.55$ $\mu\textrm{m}$, which is close to what can be obtained for the same PDMS material using the parameters in \citet{oliver2023adhesion}. For the PDMS material we have characterized, using $h_0=30.8$ nm, one obtains that the size of the process zone that fits the experimental data in Fig. \ref{fig:comparison}a is $l_{0}\simeq30.0$ nm. 

Indeed, determining the length of the process zone in viscoelastic crack propagation is still an open question. In the \citet{de1996soft} and \citet{saulnier2004adhesion} theories, the size of the "nonlinear" zone is assumed to be a constant, and the fracture energy has its maximum amplification at intermediate speeds. In PB theory this size is not constant and is directly proportional to the applied energy release rate $G$, which results in a model practically coincident with the cohesive zone model of Knauss and Schapery (see \citet{knauss2015review}). However, in fitting experimental data of fracture \citet{hui2022steady} consider two examples, a styrene-butadiene co-polymer from \citet{gent1994interfacial}, where they don't have independent estimate of the cohesive strength, but simply fit the fracture energy vs speed data, obtaining a process zone size at low speed of a nonphysical size of $0.1$ nm, consistent with \citet{gent1994interfacial}. In the second example, they consider a polyurethane elastomer called Solithane 113 of \citet{knauss2015review}, and obtain by the same process a size of $1$ nm. Hence, \citet{hui2022steady} conclude that this size cannot realistically represent a dissipation zone for which a lower bound should be the length of the monomer unit $\approx 46$ nm \citep{lake1967strength}.
\footnote{Recent literature contributions have started to question the validity of classical linear elastic fracture mechanics for unfilled plastics and elastomers suggesting that fracture initiates at a critical tensile strength, see \citet{wang2024fracture}.}
Notice that non linear crack propagation theories have been developed by Schapery using cohesive models \citep{schapery1984correspondence}, and have provided a fracture process zone at low speeds of approximately $10$ nm, much more realistic than the $0.1$ nm found by \citet{knauss2015review} and \citet{schapery1975theory2} for Solithane rubber. In fitting crack propagation data in rubbers, \citet{schapery2022stable} (Tab. 1) found a jump in propagation speed at a certain applied load which seems to suggest a sharp change of cohesive zone fracture energy as function of speed. He found a low speed fracture energy which is higher than the fast propagation speed fracture energy of a factor of about 6. In our adhesion experiments, our theory is linear and hence we cannot exclude that a non-linear theory would explain this apparent continuous change of cohesive zone fracture energy with speed of the linear theory, which is an increase with speed rather than a decrease and hence gives no instability.

\citet{barthel2024linear} reports post mortem experimental measurements of the process zone length from damage occurred at the crack tip and shows this should be of a physically reasonable size of the order of microns. Also, it clearly increases with size as PB and Schapery suggest, but contrary to the original DeGennes and Sauliner theories. Furthermore, linear theories seems to work better for very viscoelastic solids, namely when the glass transition temperature is above ambient temperature, perhaps because for very viscoelastic materials the dissipation in the bulk becomes dominant \citep{barthel2024linear}.

We have estimated the experimental effective surface energy $\Delta\gamma_{eff}(v)$ (Fig. \ref{fig:comparison}b, left y-axis) as obtained from experiments (red squares), fitted by a Gent-Schultz \citep{gent1972effect} power law equation (green dashed curve)

\begin{equation}
\Delta \gamma_{eff} = \Delta \gamma_{0} \left( 1+ \left( \frac{v}{v_0} \right)^\xi \right)\;, \label{GS}
\end{equation}
and estimated from our linear BEM numerical scheme (black solid line), respectively $\{(\Delta\gamma_{eff})_{exp}, \\(\Delta\gamma_{eff})_{GS},(\Delta\gamma_{eff})_{lin}\}$ in Fig. \ref{fig:comparison}b. To estimate the crack velocity at pull-off from the retraction rates used in the experiments we used the approximate relationship in Eq. (\ref{eq:Vvsr}), and this shows that a linear theory would fit the data much better (see dashed green line in Fig. \ref{fig:comparison}) if we assume a rate-dependent surface energy. 

Indeed, even considering that $l_{0}\simeq30.0$ nm is a more realistic estimate of the length of the fracture process zone, still we have shown that above $v=100\; \mu$m/s the linear theory largely underestimates the effective surface energy as shown in Fig. \ref{fig:comparison}b. Hence, other rate-dependent causes of dissipation seems to be at play which consistently contribute to determine the overall energy to be spent for the crack to propagate. 

As we have demonstrate numerically, linear theories such as PB theory, successfully estimate the dissipation happening within the bulk material, but they fail to account the rate-dependent nonlinear dissipative processes taking place within the process zone. Clearly, the assumption of constant intrinsic fracture energy and cohesive stress in the cohesive zone where large strain, high strain rate and non linear deformations (including damage) happen, is questionable as noticed by a very recent contribution by \citet{barthel2024linear}. Introducing the dissipative contribution coming from the nonlinear phenomena happening within the process zone, would ultimately require additional constants to be determined from actual measurements, unless one aims at describing all the nonlinear process happening within the process zone. Given the considerable
effort in the theory in characterizing the viscoelastic linear properties,
these recent models are trying mostly to understand how much of the fracture
energy amplification comes from the bulk dissipation and how much from the
cohesive zone process rate-dependence. In this respect, the estimate we gave in Fig. \ref{fig:comparison}b suggests that in our experiments at $v=10^4 \mu$m/s the nonlinear rate-dependent dissipative contribution originated within the process zone $(\Delta\gamma_{eff})_{GS}-(\Delta\gamma_{eff})_{lin}$ equals the one coming from the dissipation in the bulk $(\Delta\gamma_{eff})_{lin}$ (blue curve).

\section{Conclusions}\label{sec:conc}

We have studied the adhesive contact between a rigid Hertzian indenter and a
substrate constituted by a broad spectrum viscoelastic halfspace. For the
material we have adopted a Modified Power-Law (MPL) material model, originally
proposed by \citet{williams1964structural}, that we have extended to provide closed-form results for the creep compliance function and for the relaxation function in time domain, and also for the complex modulus and the complex compliance in the frequency domain. Notably, the MPL model is a function of only 4 parameters, the two moduli, a characteristic exponent $n$ and a characteristic time $\tau_{0}$. In particular, by changing the exponent $n$, we have shown that it is possible to have a realistic description of a broad-band viscoelastic material, which we have demonstrated by fitting the complex modulus measured for a PDMS sample. 

By using a numerical model based on the Boundary Element Method (BEM), extensive numerical studies have been performed in a wide range of the unloading rate, spanning about 8 orders of magnitude. We have shown that due to viscoelasticity, the effective surface energy can be strongly enhanced with respect to the thermodynamic surface energy, nevertheless to avoid finite size effects a certain minimum contact radius has to be reached, which we named a "threshold contact radius" $a_{0t}$. Our numerical simulations have shown that $a_{0t}$ is independent on the material exponent, but it depends on the pull-off enhancement that has to be reached at high unloading velocity. 

Provided that finite size effects are avoided $\left(a_0>a_{0t}\right)$, the theory of \citet{persson2005crack} can be used to determine the pull-off force of the spherical indenter as a function of the crack speed at pull-off with high accuracy, but only within the assumptions of the linear theory and rate-independent fracture process zone parameters. Relating the numerical results based on a Lennard-Jones force-separation law to the theory of \citet{persson2005crack} required to define a parameter $\alpha=0.3491$ of order unity that relates the critical stress $\sigma_c$ in PB theory to the maximum stress used in the LJ law $\sigma_0$, which was found independent on the ratio $k=E_0/E_\infty$. Adhesion experiments are usually run in displacement control, and the crack speed at pull-off is certainly not a control parameter, nevertheless we have shown that the velocity of the crack at pull-off $\widehat{v}$ scales as $\widehat{v}\simeq2.887\widehat{r}_{PB}^{1.171}$ over more then 8 orders of magnitude, which provides an extremely simple relation to roughly estimate the pull-off force starting only from the material model parameters and the unloading rate with about $\pm15 \%$ confidence.

Finally, by using the MPL for the viscoelastic material and the developed BEM code, we have attempted a comparison between the numerical and the experimental results, which turned out to be satisfactorily accurate up to unloading rates $r=100$ $\mu$m/s, while for faster
unloading the numerical results predict lower enhancement with respect to
what is measured by our experiments. This observation turns out to be in good agreement with previous Literature results Refs. \citep{vandonselaar2023silicone,tiwari2017effect}, where similar experiments were conducted. 

A non linear description of the material behaviour must be necessarily a better description than linear, so perhaps the $J$ integral approach of Schapery \citep{schapery2023crack} could improve our results. However, as in classical non linear fracture mechanics, we ultimately need to measure experimentally the critical value of the fracture energy, which cannot be found reliably from other material properties, in viscoelastic adhesion even if some progress is made by the linear theories, the estimate of the bulk dissipation contribution to fracture energy enhancement is not sufficient, and, ultimately,
the fracture process zone rate-dependency must be measured experimentally.
Hence, at present, the measurement of the $\Delta\gamma_{eff}(v)$ curve remains the only engineering approach, resulting in the phenomenological \citet{gent1972effect} law. Notice that if we use the measured Gent-Schultz law with a power $\xi=0.41$ and assume the far field material is elastic with relaxed modulus, we can solve the adhesive contact problem using the Muller solution as corrected in \citet{ciavarella2021improved}. This results in a pull-off force which doesn't scale with the same power law  of the Gent-Schulz law, but with power $0.27$ in this case, so also the Muller solution is 
misleading. 

\appendix
\section{Modified power law material model}
\label{sec:appendixA}

\subsection{Relaxation function in time and frequency domain}

Let us assume to model a viscoelastic material with a continuous distribution
$H\left(  \tau\right)  $ of relaxation times, which is the so-called material
relaxation spectrum, in parallel with a Hookean spring giving the material
stiffness for long time. This coincides with assuming a Wiechert model (see Fig. \ref{fig:A1}) with an
infinite number of Maxwell arms. The general relation for the stress
$\sigma\left(  t\right)  $ at time $t$ is (Eq. (2.34) in \citet{williams1964structural})\footnote{Notice that we are using the notation according to \citet{williams1964structural}. In \citet{christensen2012theory} book the relaxation spectrum is defined as $[H(\tau)]_{Christensen}=[H(\tau)]_{Williams}/\tau$}:%

\begin{equation}
\sigma\left(  t\right)  =\left\{  E_{0}+\int_{0}^{\infty}\frac{H\left(
\tau\right)  }{\left[  \frac{d}{dt}+1/\tau\right]  \tau}d\tau\frac{d}%
{dt}\right\}  \varepsilon\left(  t\right)\;,  \label{sigt}%
\end{equation}
Converting Eq.
(\ref{sigt}) in the frequency domain, we get:%

\begin{equation}
{\sigma}\left(  \omega\right)  =\left\{  E_{0}+\int_{0}^{\infty}%
\frac{H\left(  \tau\right)  \boldsymbol{i}\omega}{\left[  \boldsymbol{i}%
\omega\tau+1\right]  }d\tau\right\}  {\varepsilon}\left(
\omega\right)  =\overline{E}\left(  \omega\right)  {\varepsilon}\left(
\omega\right)\;, 
\end{equation}
where, $\boldsymbol{i}$ is the imaginary unit, $\omega$ is the angular
frequency, and, by definition, $\overline{E}\left(  \omega\right)  =E^{\prime
}\left(  \omega\right)  +\boldsymbol{i}E^{\prime\prime}\left(  \omega\right)
$ is the complex modulus, hence:%
\begin{equation}
\overline{E}\left(  \omega\right)  =E_{0}+\int_{0}^{\infty}\frac{H\left(
\tau\right)  \boldsymbol{i}\omega}{\left[  \boldsymbol{i}\omega\tau+1\right]
}d\tau\label{Ecmpl}\;.%
\end{equation}

In order to fit the experimental data, one can guess a certain form for the
relaxation spectrum $H\left(  \tau\right)$. As suggested by \citet{williams1964structural},
a broad-band approximation of the response of the viscoelastic material can be
obtained by adopting for the relaxation spectrum a modified power law:

\begin{equation}
H\left(  \tau\right)  =\left(  \frac{E_{\infty}-E_{0}}{\Gamma\left(  n\right)
}\right)  \left(  \frac{\tau_{0}}{\tau}\right)  ^{n}\exp\left(  -\frac
{\tau_{0}}{\tau}\right)\;,  \label{app:relspec}
\end{equation}

The complex modulus is $\overline{E}\left(  \omega\right)  =E^{\prime}\left(
\omega\right)  +\boldsymbol{i}E^{\prime\prime}\left(  \omega\right)  $
can be written in terms of the relaxation spectrum:
\begin{align}
\overline{E}\left(  \omega\right)    & =E_{0}+\int_{0}^{\infty}\frac{H\left(
\tau\right)  \boldsymbol{i}\omega}{\left[  \boldsymbol{i}\omega\tau+1\right]
}d\tau\;,\\
E^{\prime}\left(  \omega\right)    & =E_{0}+\int_{0}^{\infty}\frac{H\left(
\tau\right)  \omega^{2}\tau}{\left[  1+\omega^{2}\tau^{2}\right]  }%
d\tau\label{Eprime}\;,\\
E^{\prime\prime}\left(  \omega\right)    & =\int_{0}^{\infty}\frac{H\left(
\tau\right)  \omega}{\left[  1+\omega^{2}\tau^{2}\right]  }d\tau\;.
\label{Esecond}%
\end{align}
By using Eq. (\ref{app:relspec}) for the relaxation spectrum $H\left(
\tau\right)  $ one obtains

\begin{align}
\overline{E}\left(  \omega\right) & =E_{0}+\left(  E_{\infty}-E_{0}\right)
\boldsymbol{i}\omega\tau_{0}\exp\left(  \boldsymbol{i}\omega\tau_{0}\right)
\mathbf{E}_{n}\left(  \boldsymbol{i}\omega\tau_{0}\right)\;,\\
E^{\prime}\left(  \omega\right)   & =E_{0}+\frac{\left(  E_{\infty}%
-E_{0}\right)  }{\Gamma\left(  n\right)  }\left\{
\begin{array}
[c]{c}%
\pi\left(  \tau_{0}\omega\right)  ^{n}\cos\left(  n\frac{\pi}{2}+\tau
_{0}\omega\right)  \csc\left(  n\pi\right)  +...\\
...+\left(  \tau_{0}\omega\right)  ^{2}\Gamma\left(  n-2\right)  _\text{p}\text{F}_\text{q}
\left[  1;\left\{  \frac{3-n}{2},2-\frac{n}{2}\right\}  ;-\frac{\left(
\tau_{0}\omega\right)  ^{2}}{4}\right]
\end{array}
\right\} \;,\\
E^{\prime\prime}\left(  \omega\right)   & =\frac{\left(  E_{\infty}%
-E_{0}\right)  }{\Gamma\left(  n\right)  }\left\{
\begin{array}
[c]{c}%
\pi\left(  \tau_{0}\omega\right)  ^{n}\sin\left(  n\frac{\pi}{2}+\tau
_{0}\omega\right)  \csc\left(  n\pi\right)  +...\\
...+\left(  \tau_{0}\omega\right)  \Gamma\left(  n-1\right)  _\text{p}\text{F}_\text{q}\left[
1;\left\{  1-\frac{n}{2},\frac{3-n}{2}\right\}  ;-\frac{\left(  \tau_{0}%
\omega\right)  ^{2}}{4}\right]
\end{array}
\right\}\;,
\end{align}
where $\tau_{0}$ is the characteristic time, $n>0$ is a characteristic
exponent, $\omega$ is the angular frequency, $E_{0}$ is the relaxed elastic
modulus, $E_{\infty}$ is the instantaneous elastic modulus, $_\text{p}\text{F}_\text{q}[a;b;z]$
is the generalized hypergeometric function, $\Gamma\left(  x\right)  $ is the
Euler gamma function.

\begin{figure}[t]
\centering
\includegraphics[width=4in]{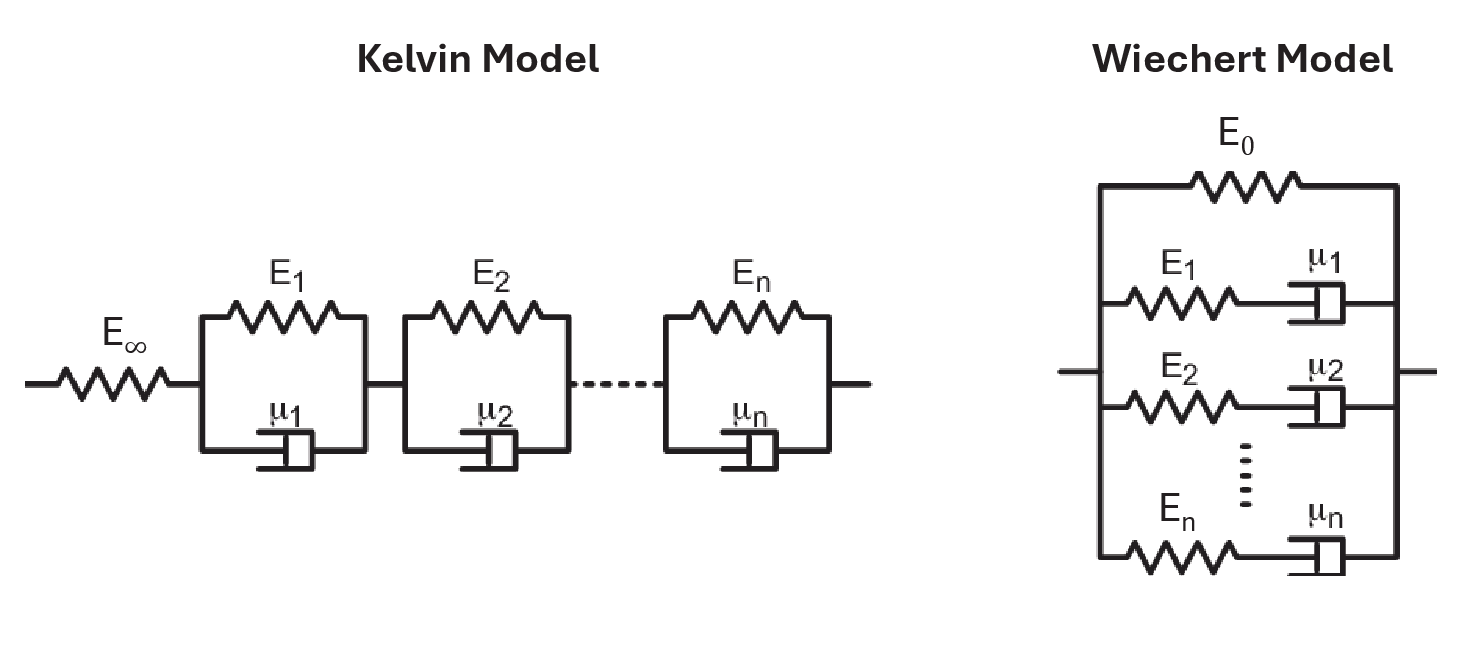}
\caption{The kelvin model (left) and the Wiechert model (right) for the representation of the mechanical behaviour of a viscoelastic material.} \label{fig:A1}
\end{figure}

\subsection{Compliance function in time and frequency domain}
Let us now consider to model a viscoelastic material with an infinite series of Voigt elements, in series with a Hookean spring giving the material stiffness for
short time, the so-called Kelvin model (see Fig. \ref{fig:A1}). This coincides with assuming a continuous distribution of retardation times $L\left(  \tau\right) $, which is the so-called material retardation
spectrum. A general relation for the deformation $\varepsilon\left(
t\right) $ at time $t$ is given by (Eq. (2.42) in \citet{williams1964structural}):

\begin{equation}
\varepsilon\left(  t\right)  =\left\{  C_{\infty}+\int_{0}^{\infty}%
\frac{L\left(  \tau\right)  }{\left[  \frac{d}{dt}+1/\tau\right]  \tau^{2}%
}d\tau\right\}  \sigma\left(  t\right)\;,  \label{epst}%
\end{equation}
where $C_{\infty}=1/E_{\infty}$ is the creep compliance in the glassy limit.
Converting Eq. (\ref{epst}) in the frequency domain gives:

\begin{equation}
{\varepsilon}\left(  \omega\right)  =\left\{  C_{\infty}+\int
_{0}^{\infty}\frac{L\left(  \tau\right)  }{\left[  \boldsymbol{i}\omega
+1/\tau\right]  \tau^{2}}d\tau\right\}  {\sigma}\left(
\omega\right)  = \overline{C}\left(  \omega\right)  {\sigma}\left(
\omega\right)\;,
\end{equation}
where, $\boldsymbol{i}$ is the imaginary unit and, by definition, $\overline{C}\left(  \omega\right)  =C^{\prime}\left(  \omega\right)  -\boldsymbol{i}%
C^{\prime\prime}\left(  \omega\right)$ is the complex compliance. Hence, we have:%

\begin{equation}
\overline{C}\left(  \omega\right)  =C_{\infty}+\int_{0}^{\infty}\frac{L\left(
\tau\right)  }{\left[  \boldsymbol{i}\omega+1/\tau\right]  }\frac{d\tau}%
{\tau^{2}}\;. \label{Dcmpl}%
\end{equation}

We note that to match the experimental data, a specific form for the retardation spectrum $L\left(  \tau\right)$ could be considered.  Following \citet{williams1964structural} suggestion, a broad-band approximation of the viscoelastic material response can be achieved by using a modified power law for the retardation spectrum, such as:
\begin{equation}
L\left(  \tau\right)  =\left(  \frac{C_{0}-C_{\infty}}{\Gamma\left(  n\right)
}\right)  \left(  \frac{\tau}{\tau_{0}}\right)  ^{n}\exp\left(  -\frac{\tau
}{\tau_{0}}\right)\;,  \label{app:retspec}%
\end{equation}

The complex compliance is defined as follows:
\begin{equation}
\overline{C}\left(  \omega\right)
=C^{\prime}\left(  \omega\right)  -\boldsymbol{i}C^{\prime\prime}\left(
\omega\right)\,,
\end{equation}
where
\begin{align}
\overline{C}\left(  \omega\right)    & =C_{\infty}+\int_{0}^{\infty}\frac{L\left(
\tau\right)  }{\left[  \boldsymbol{i}\omega+1/\tau\right]  }\frac{d\tau}%
{\tau^{2}}\;,\\
C^{\prime}\left(  \omega\right)    & =C_{\infty}+\int_{0}^{\infty}%
\frac{L\left(  \tau\right)  }{\left[  1+\omega^{2}\tau^{2}\right]  }%
\frac{d\tau}{\tau}\label{Dprimeomega}\;,\\
C^{\prime\prime}\left(  \omega\right)    & =\int_{0}^{\infty}\frac{L\left(
\tau\right)  \omega}{\left[  1+\omega^{2}\tau^{2}\right]  }d\tau\;.
\label{Dsecomega}%
\end{align}
By using Eq. (\ref{app:retspec}) for the retardation spectrum $L\left(
\tau\right)  $ one obtains

\begin{align}
\overline{C}\left(  \omega\right) & =C_{\infty}+\frac{\left(  C_{0}-C_{\infty
}\right)  }{\boldsymbol{i}\omega\tau_{0}}\exp\left(  -\frac{\boldsymbol{i}%
}{\omega\tau_{0}}\right)  \mathbf{E}_{n}\left(  -\frac{\boldsymbol{i}}%
{\omega\tau_{0}}\right)\;\\
C^{\prime}\left(  \omega\right)   & =C_{\infty}+\frac{\left(  C_{0}-C_{\infty
}\right)  \left(  \tau_{0}\omega\right)  ^{-2-n}}{\Gamma\left(  n\right)
}\left\{
\begin{array}
[c]{c}%
\pi\left(  \tau_{0}\omega\right)  ^{2}\cos\left(  n\frac{\pi}{2}+\frac{1}%
{\tau_{0}\omega}\right)  \csc\left(  n\pi\right)  +...\\
...+\left(  \tau_{0}\omega\right)  ^{n}\Gamma\left(  -2+n\right)  _\text{p}\text{F}_\text{q}
\left[  1;\left\{  \frac{3-n}{2},2-\frac{n}{2}\right\}  ;-\frac{1}{4\left(
\tau_{0}\omega\right)  ^{2}}\right]
\end{array}
\right\} \\
C^{\prime\prime}\left(  \omega\right)   & =\frac{\left(  C_{0}-C_{\infty
}\right)  \left(  \tau_{0}\omega\right)  ^{-1-n}}{\Gamma\left(  n\right)
}\left\{
\begin{array}
[c]{c}%
\pi\tau_{0}\omega\sin\left(  n\frac{\pi}{2}+\frac{1}{\tau_{0}\omega}\right)
\csc\left(  n\pi\right)  +...\\
...+\left(  \tau_{0}\omega\right)  ^{n}\Gamma\left(  -1+n\right)  _\text{p}\text{F}_\text{q}
\left[  1;\left\{  1-\frac{n}{2},\frac{3-n}{2}\right\}  ;-\frac{1}{4\left(
\tau_{0}\omega\right)  ^{2}}\right]
\end{array}
\right\}\;,
\end{align}
where $\tau_{0}$ is the characteristic time, $n>0$ is a characteristic
exponent, $\omega$ is the angular frequency, $C_{0}=1/E_{0}$ is the relaxed
compliance, $C_{\infty}=1/E_{\infty}$ is the instantaneous compliance,
$_\text{p}\text{F}_\text{q}[a;b;z]$ is the generalized hypergeometric function, $\Gamma\left(
x\right)  $ is the Euler gamma function.

\section{Details of the PB model for the effective surface energy}
\label{sec:AppendixB}

According to PB theory the dimensionless effective surface energy for a MPL viscoelastic material model $\widehat{\Gamma}_{eff}$ is obtained as 
\begin{equation}
\widehat{\Gamma}_{eff}=\left[  1-\left(  1-k\right)  I\left(  n,\widehat{v},\widehat{\Gamma}_{eff}\right)
\right]  ^{-1}\,, 
\end{equation}
where $I(n,\widehat{v},\widehat{\Gamma}_{eff})$ stands for the integral in Eq. (\ref{PBmodel}), which can be evaluated in closed form as:
\begin{align}
I(n,\widehat{v},\widehat{\Gamma}_{\text{eff}})   &  =\frac{2^{(-3-2n)}\pi^{(-3/2-n)}
}{(n-1)\widehat{v}} \biggl\{-4^{(1+n)}\pi^{(1/2+n)}\biggl[\widehat{\Gamma}_{\text{eff}} +\\
& -2(n-1)\pi \widehat{v} _\text{p}\text{F}_\text{q} \biggl[{-}\frac{{1}}{2}; \left\{  \frac{{1}}{2}{-}\frac{{n}}{2}{,1-}%
\frac{{n}}{2}\right\}  ;-\left(  \frac{\widehat{\Gamma}_{\text{eff}}}{4\pi \widehat{v}}\right)
^{2}\biggr]\biggr]  +  \nonumber \\
& +2\pi\left(  \frac{\widehat{\Gamma}_{\text{eff}}}{\widehat{v}}\right)  ^{n}\left[  4\pi \widehat{v}\Gamma\left[
1-\frac{n}{2}\right]  \left.  _\text{p}\text{F}_\text{q}\right\vert _{\text{Reg}}\left[
\frac{{n-1}}{2};\left\{  \frac{{1}}{2}{,}\frac{{2+n}}{2}\right\}  ;-\left(
\frac{\widehat{\Gamma}_{\text{eff}}}{4\pi \widehat{v}}\right)  ^{2}\right]  +\right.  \nonumber\\
& +\widehat{\Gamma}_{\text{eff}}\Gamma\left[  \frac{3}{2}-\frac{n}{2}\right]
\left.  _\text{p}\text{F}_\text{q}\right\vert _{\text{Reg}}\biggl[  \frac{{n}}{2};\left\{
\frac{{3}}{2}{,}\frac{{(3+n)}}{2}\right\}  ;-\left(  \frac{\widehat{\Gamma}_{\text{eff}}}{4\pi
\widehat{v}}\right)  ^{2}\biggr]  \biggr]  \biggr\}\;,
\end{align}

where $_\text{p}\text{F}_\text{q}[{a},b,z]$ is the generalized hypergeometric function,
$\left.  _\text{p}\text{F}_\text{q}\right\vert _{\text{Reg}}[{a},b,z]$ is the regularized
generalized hypergeometric function and $\Gamma\left[  x\right]  $ is the gamma function (we used Wolfram Mathematica$^\text{\textcopyright}$ for algebraic manipulation).

To ease the use of Eq. (\ref{PBmodel2}) we report here in Fig. (\ref{fig:B1}) the quantity $\widehat{\Gamma}_{eff}-1$, showing that there exists two power law regimes, the first is a linear scaling where $(\widehat{\Gamma}_{eff}-1)\simeq q_1 V^1$, with the coefficient $q_1=-0.0125+3.136n$ depending on the material exponent "$n$", the second instead can be written as $(\widehat{\Gamma}_{eff}-1)\simeq q_2 V^{m_2}$, with the $\{q_2,m_2\}$ constants depending on $n$, which is shown in the inset of Fig. \ref{fig:B1}. Notice that, a SLS would have $n\approx1.6$ which provide a scaling of $(\widehat{\Gamma}_{eff}-1)\propto V^{0.5}$, while broad band materials provide a much lower exponent $m_2$ as one can see in Fig. \ref{fig:B1} (inset).
\begin{figure}[h!]
\centering
\includegraphics[width=4in]{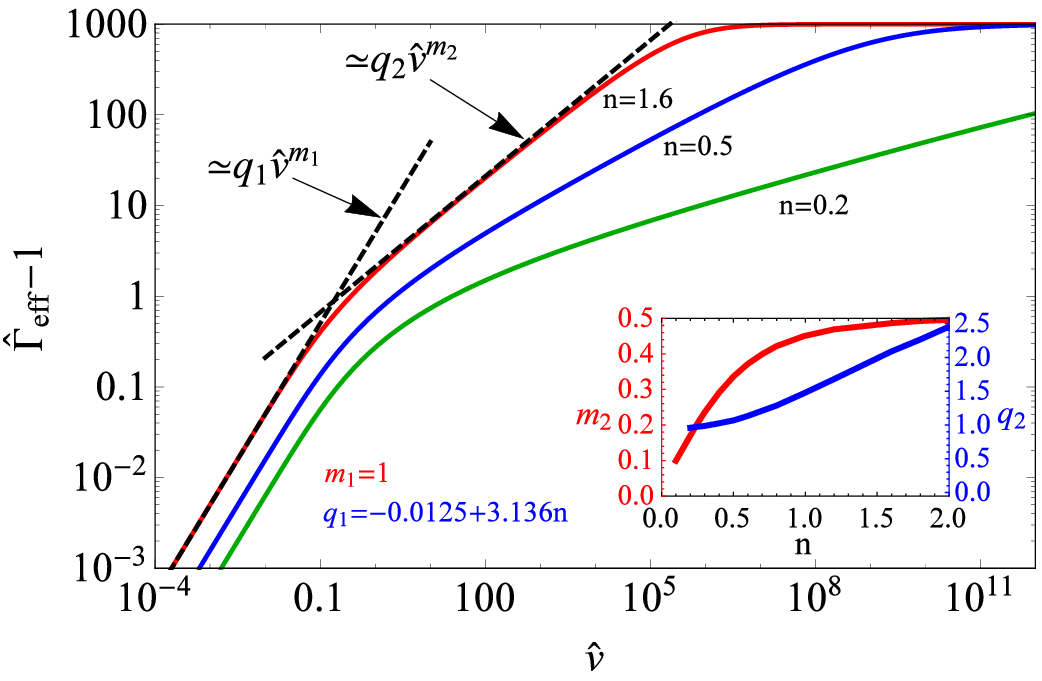}
\caption{Fit of the enhancement of the effective surface energy $\widehat{\Gamma}_{eff}-1$ obtained using a MPL material model in PB theory. Two power law scaling have been identified for $\widehat{v}<1$ and for $\widehat{v}>1$, which coefficients are given in the figure and in the inset as a function of the material exponent $n$.} \label{fig:B1}
\end{figure}

\section*{Acknowledgment}
A.M., M.T., M.C., A.P. were partly supported by the Italian Ministry of University and Research under the Programme “Department of Excellence” Legge 232/2016 (Grant No. CUP - D93C23000100001). A.P., A.M. and M.T. were supported by the European Union (ERC-2021-STG, “Towards Future Interfaces With Tuneable Adhesion By Dynamic Excitation” - SURFACE, Project ID: 101039198, CUP: D95F22000430006). Views and opinions expressed are however those of the authors only and do not necessarily reflect those of the European Union or the European Research Council. Neither the European Union nor the granting authority can be held responsible for them. A.P. was partly supported by the European Union through the program – Next Generation EU (PRIN-2022-PNRR, "Fighting blindness with two photon polymerization of wet adhesive, biomimetic scaffolds for neurosensory REtina-retinal Pigment epitheliAl Interface Regeneration" - REPAIR, Project ID: P2022TTZZF, CUP: D53D23018570001). Q.W. was supported by National Natural Science Foundation of China (No. 12025203).  

\section*{Data availability}
The dataset generated for this article will be available on Zenodo.

\bibliographystyle{elsarticle-harv} 
\bibliography{cas-refs}
\end{document}